\documentclass[11pt,prd,superscriptaddress,preprintnumbers,amsmath,amssymb,nofootinbib]{revtex4}
\usepackage{epsfig}  
\usepackage{graphicx}
\usepackage{hyperref}
\hypersetup{
    colorlinks=true,
    citecolor=red,
    linkcolor=blue,
    filecolor=green,      
    urlcolor=magenta,
}
\usepackage{color}
\usepackage{float}
\usepackage{amsfonts}
\usepackage{amsmath}
\usepackage{slashed}
\usepackage{soul}
\large

\begin{document}

\title{Signatures of complex new physics in $b\to c\tau\bar{\nu}$ transitions}

\author{Suman Kumbhakar\footnote{Presently at Centre for High Energy Physics, Indian Institute of Science Bangalore 560012}}
\email{suman@phy.iitb.ac.in}
\affiliation{Indian Institute of Technology Bombay, Mumbai 400076, India}

\begin{abstract}
The anomalies in the measurements of $R_D$ and $R_{D^*}$ continue to provide motivation for physics beyond the Standard
Model. In this work, we assume the new physics Wilson coefficients to be complex and find their values by doing a global fit to the present $b\rightarrow c\tau\bar{\nu}$ data. We find that the number of allowed solutions depend on the choice of the upper limit on $Br(B_c\rightarrow \tau\bar{\nu})$. We find that the forward-backward asymmetries in $B\rightarrow (D, D^*)\tau\bar{\nu}$ decays have the capability to distinguish between different solutions. Further we calculate the maximum values of CP violating triple product asymmetries in $B\to D^*\tau\bar{\nu}$ decay allowed the current data. We observe that only one of the three CP asymmetries can be enhanced up to a maximum value of $\sim 2-3\%$ whereas the other asymmetries remain smaller.
\end{abstract}
 
\maketitle 

\section{Introduction} 

The heavy meson decays, in particular the $B$ meson decays, are a very fertile ground to probe possible physics beyond the
Standard Model (SM). In the past few years, several measurements by BaBar, Belle and LHCb in the $B$ meson decays show significant deviations from their SM predictions. One such class of decays occurs through the charged current $b\rightarrow c\tau\bar{\nu}$ transition which is a tree level process in the SM. In this sector, two interesting observables are
\begin{equation}
R_{D} = \frac{\mathcal{B}(B\rightarrow D\tau\bar{\nu})}{\mathcal{B}(B\rightarrow D \{e/\mu\}  \bar{\nu})}\,,\quad
R_{D^{*}} = \frac{\mathcal{B}(B\rightarrow D^{*}\tau\bar{\nu})}{ \mathcal{B}(B\rightarrow D^{*} \{e/\mu\} \bar{\nu})}\,.
\end{equation}
These flavor ratios are consecutively measured by BaBar \cite{Lees:2012xj,Lees:2013uzd}, Belle \cite{Huschle:2015rga,Sato:2016svk,Hirose:2016wfn,Abdesselam:2019dgh} and LHCb \cite{Aaij:2015yra,Aaij:2017uff,Aaij:2017deq} collaborations. The SM predicts $R_D$ to be $0.299\pm 0.003$ whereas the present experimental world average is $0.340\pm 0.027\,(\rm stat.)\pm 0.013\,(\rm syst.)$. For $R_{D^*}$, the SM prediction is $0.258\pm 0.005$ and the experimental world average is $0.295\pm 0.011\,(\rm stat.)\pm 0.008\,(\rm syst.)$. The SM predictions and the world averages are noted down from Heavy Flavor Averaging Group~\cite{Amhis:2019ckw}. 
The present average values of $R_{D}$ and $R_{D^*}$ exceed the SM predictions by $1.4\sigma$ and $2.5\sigma$ respectively. Including the correlation of $-0.38$, the tension between the measurements and the SM predictions is at the level of $3.1\sigma$. This discrepancy is an indication of lepton flavor universality (LFU) violation between $\tau$ and $\mu/e$ leptons.

In addition, the LHCb collaboration measured another flavor ratio
$R_{J/\psi} = \Gamma(B_c\rightarrow J/\psi  \tau  \bar{\nu})/\Gamma(B_c\rightarrow J/\psi  \mu \bar{\nu})$ 
whose value is $0.71\pm 0.17\,(\rm stat.) \pm 0.18\,(\rm syst.)$~\cite{Aaij:2017tyk}. Eventhough the uncertainties are quite large, it is $1.7\sigma$ higher than its SM prediction $0.289\pm 0.010$ \cite{Dutta:2017xmj}. This is an additional hint of LFU violation in the $b\rightarrow c \ell  \bar{\nu}$ sector. These deviations could be due to presence of new physics (NP) either in $b\rightarrow c\tau\bar{\nu}$ or in $b\rightarrow c\{\mu,e\}\bar{\nu}$ transition. However, it has been shown in Refs.~\cite{Alok:2017qsi,Iguro:2020cpg} that the latter possibility is ruled out by other measurements. Therefore, we assume the presence of NP only in $b\rightarrow c\tau\bar{\nu}$ transition.

Apart from these, Belle collaboration has measured two angular observables in the $B \to D^* \tau \bar{\nu}$ decay $-$ (a) the $\tau$ polarization $P^{D^*}_{\tau}$ and (b) the $D^*$ longitudinal polarization fraction $F^{D^*}_L$. The measured values of these two quantities are~\cite{Hirose:2016wfn,Abdesselam:2019wbt}
\begin{eqnarray}
P^{D^*}_{\tau} &=&  - 0.38 \pm 0.51\, (\rm stat.) ^{+0.21}_{-0.16}\, (\rm syst.),\\
F^{D^*}_{L} &=& 0.60 \pm 0.08\, (\rm stat.)\pm 0.04\, (\rm syst.).
\end{eqnarray}
The measured value of $P_{\tau}^{D^*}$ is consistent with its SM prediction of $-0.497\pm0.013$ \cite{Tanaka:2012nw} whereas for $F^{D^*}_L$ it is $1.6\sigma$ higher than the SM prediction of $0.46\pm 0.04$~\cite{Alok:2016qyh}.

The anomalies in $b\rightarrow c\tau\bar{\nu}$ transition have been studied in various model independent techniques~\cite{Jung:2018lfu,Bhattacharya:2018kig,Hu:2018veh,Alok:2019uqc,Asadi:2019xrc,Murgui:2019czp,Bardhan:2019ljo,Blanke:2019qrx,Shi:2019gxi,Becirevic:2019tpx,Sahoo:2019hbu,Cheung:2020sbq,Cardozo:2020uol}. The Wilson coefficients (WCs) of the 
NP operators are determined by doing a fit to the data available in this sector along with the constraint on the branching ratio of $B_c\to \tau \bar{\nu}$ decay. In Ref.~\cite{Alok:2019uqc}, it has been shown that the NP Lorentz structure in form of $(V-A)\times (V-A)$ is the only one operator solution allowed by the present data.

In this paper we do a global fit to of all present data on $b\rightarrow c\tau\bar{\nu}$ transition by starting with a most general effective Hamiltonian. Assuming the NP WCs to be complex, we find the allowed NP solutions with their corresponding WCs. We show that one/two/three NP solution(s) is (are) allowed if we make the three different choices on the upper limits of $10\%$/$30\%$/$60\%$ on the branching ratio of $B_c\rightarrow \tau\bar{\nu}$. We calculate the predictions of angular observables in $B\rightarrow (D,D^*)\tau\bar{\nu}$ decays and comment on their ability to distinguish between the allowed solutions. Further, we compute the predictions of the CP violating triple product asymmetries in $B\to D^*\tau\bar{\nu}$ decay for the three NP solutions. We show that only one of these three asymmetries can be enhanced up to a maximum of $\sim 2-3\%$ in the presence the allowed NP scenarios.
 
The paper is organized as follows. In Section II, we describe our methodology for calculation and present our fit results. In this section, we calculate the predictions of the angular observables of $B\rightarrow (D,D^*)\tau\bar{\nu}$ decays and discuss their distinguishing capabilities. In section III, we determine the maximum possible CP violating triple product asymmetries in $B\to D^*\tau\bar{\nu}$ decay allowed by the current data. We present our conclusions in section IV.

\section{Fit Methodology and Results}

We start with the most general effective Hamiltonian for $b\rightarrow c\tau\bar{\nu}$ transition which contains all possible Lorentz structures. This is expressed as \cite{Freytsis:2015qca}
\begin{equation}
\mathcal{H}_{\rm eff}= \frac{4 G_F}{\sqrt{2}} V_{cb}\left[\mathcal{O}_{V_L} + \frac{\sqrt{2}}{4 G_F V_{cb}} \frac{1}{\Lambda^2} \left\lbrace \sum_i \left(C_i \mathcal{O}_i +
 C^{'}_i \mathcal{O}^{'}_i + C^{''}_i \mathcal{O}^{''}_i \right) \right\rbrace \right],
\label{effH}
\end{equation}
where $G_F$ is the Fermi coupling constant and $V_{cb}$ is the Cabibbo-Kobayashi-Maskawa (CKM) 
matrix element. Here we assume that the neutrino is left chiral. We also assume the new physics scale $\Lambda = 1$ TeV. The five unprimed operators
\begin{equation}
\mathcal{O}_{V_L} =(\bar c \gamma_{\mu} P_L b)(\bar \tau \gamma^{\mu}  P_L \nu) \ , \quad  
\mathcal{O}_{V_R}=(\bar c \gamma_{\mu}  P_R b)(\bar \tau \gamma^{\mu}  P_L \nu) \nonumber\ , \quad
\end{equation}  
\begin{equation}
\mathcal{O}_{S_L}=(\bar c P_L b)(\bar \tau P_L \nu ), \quad
\mathcal{O}_{S_R}=(\bar c P_R b)(\bar \tau P_L \nu),  \quad    
\mathcal{O}_T=(\bar c \sigma_{\mu \nu}P_L b)(\bar \tau \sigma^{\mu \nu} P_L \nu) \ , 
\label{ops}
\end{equation}
form the complete set of operators consistent with global baryon number and lepton number conservation.  The primed
and double primed operators $\mathcal{O}^{'}_i$ and $\mathcal{O}^{''}_i$ only arise in different Leptoquark models~\cite{Freytsis:2015qca} depending on their spin and charge. A more rigorous discussion on all possible Leptoquarks can be found in Ref.~\cite{Dorsner:2016wpm}.  The Lorentz structures of all these operators are described in Ref.~\cite{Freytsis:2015qca}. In particular, $\mathcal{O}^{'}_i$ and $\mathcal{O}^{''}_i$ operators can be expressed in terms of five unprimed operators using Fierz identities. The constants $C_i$, $C^{'}_i$ and $C^{''}_i$ are the respective WCs of the NP operators in which NP effects are hidden. 
In this analysis, we assume these NP WCs to be complex.
 
Using the effective Hamiltonian given in Eq.~(\ref{effH}), we calculate the expressions of measured observables $R_D$, $R_{D^*}$, $R_{J/\psi}$, $P^{D^*}_{\tau}$ and $F^{D^*}_L$ as functions of the NP WCs. To obtain the values of NP WCs, we do a fit of these expressions to the measured values of the observables. In doing the fit, we take only one NP operator at a time. We define the $\chi^2$ function as follows
\begin{equation}
\chi^2(C_i)=\sum_{R_D, R_{D^*}, R_{J/\psi}, P_{\tau}^{D^*}, F^{D^*}_L}\left(O^{th}(C_i)-O^{exp}\right) \mathcal{C}^{-1} \left(O^{th}(C_i)-O^{exp}\right),
\label{chi2}
\end{equation}
where $O^{th}(C_i)$ are NP predictions of each observable and $O^{exp}$ are the corresponding experimental central values. The $\mathcal{C}$ denotes the covariance matrix which includes both theory and experimental correlations.

The  $B\rightarrow (D, D^{*}) \ell  \bar{\nu}$ decay distributions depend upon hadronic form-factors. The determination of these form-factors can be calculated with the HQET techniques which are presently known at $O(1/m_b, 1/m^2_c, \alpha_s)$.  In this work we use the HQET form factors in the form parametrized by Caprini {\it et al.} \cite{Caprini:1997mu}. The parameters for $B\rightarrow D$ decay are determined from the lattice  QCD \cite{Aoki:2019cca} calculations and we use them in our analyses. For $B\rightarrow D^*$ decay, the HQET parameters are extracted using data from Belle and BaBar experiments along with the inputs from lattice. In this work, the numerical values of these parameters are taken from refs. \cite{Bailey:2014tva} and  \cite{Amhis:2019ckw}. The form factors for $B_c\rightarrow J/\psi$ transition and their uncertainties from ref.~\cite{Wen-Fei:2013uea} are used in the calculation of $R_{J/\psi}^{th}$. These form factors are calculated in perturbative QCD framework.
 
\begin{figure*}[ht] 
\centering
\begin{tabular}{cc}
\includegraphics[width=75mm]{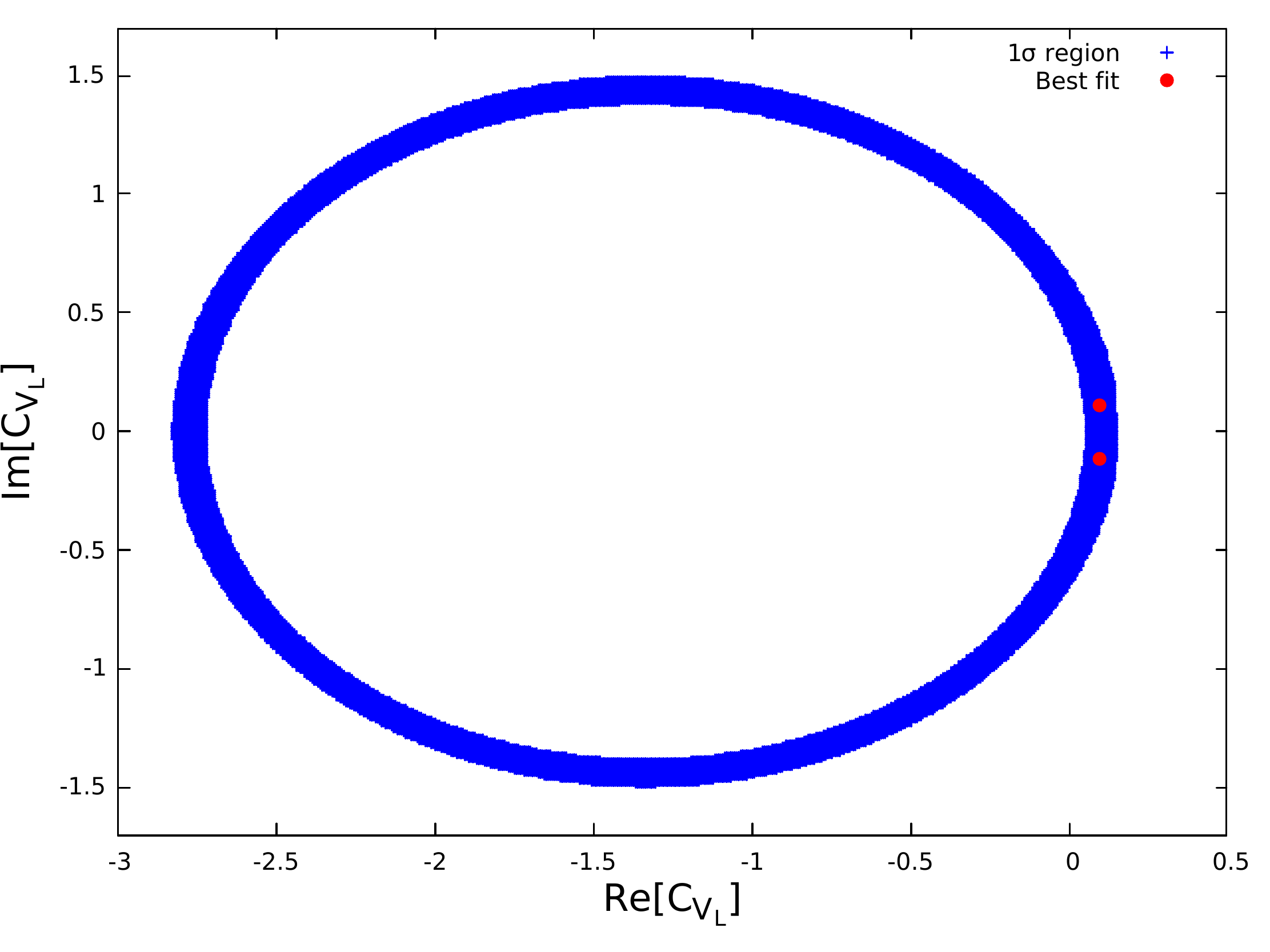}&
\includegraphics[width=75mm]{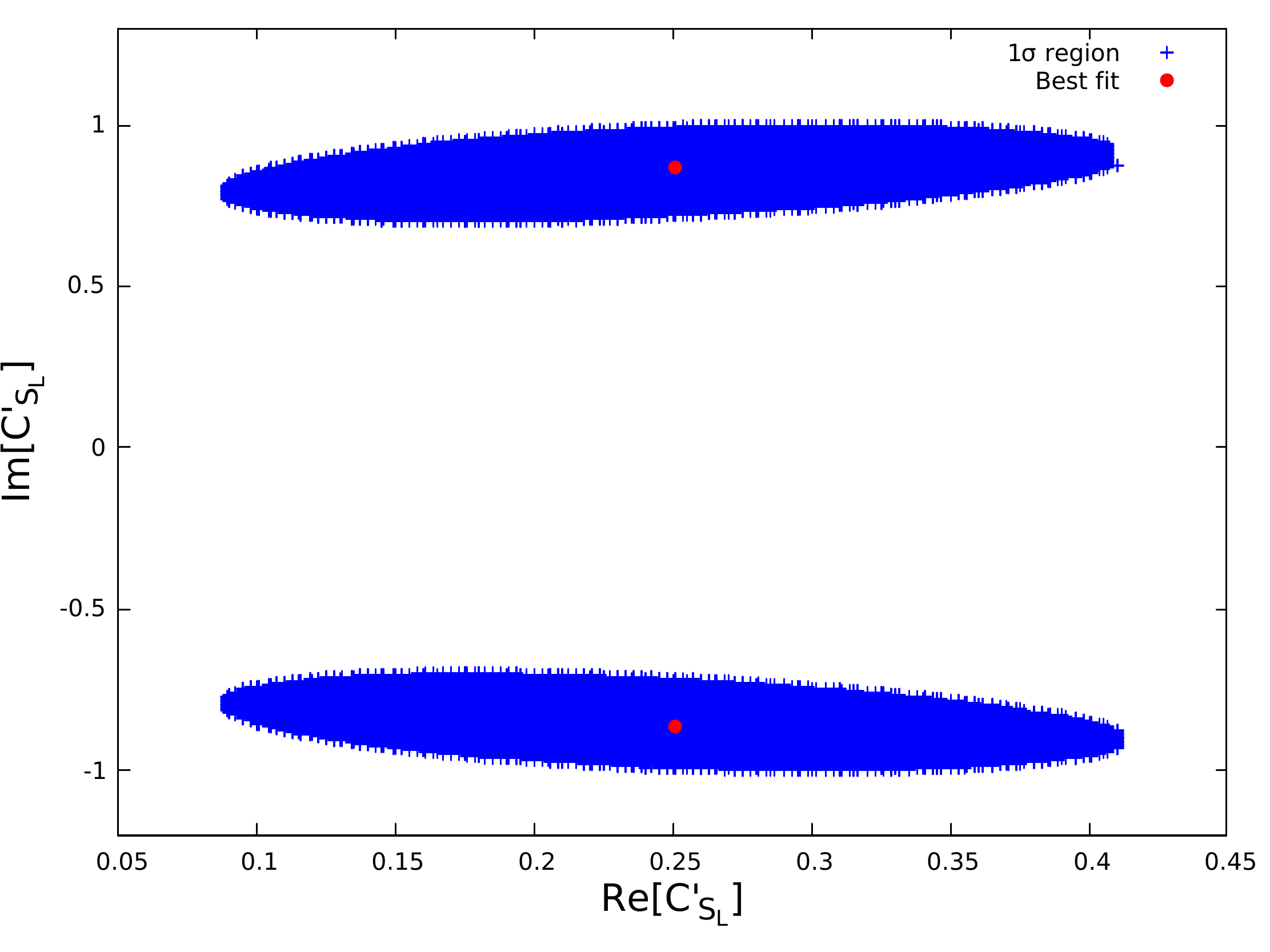}\\
\includegraphics[width=75mm]{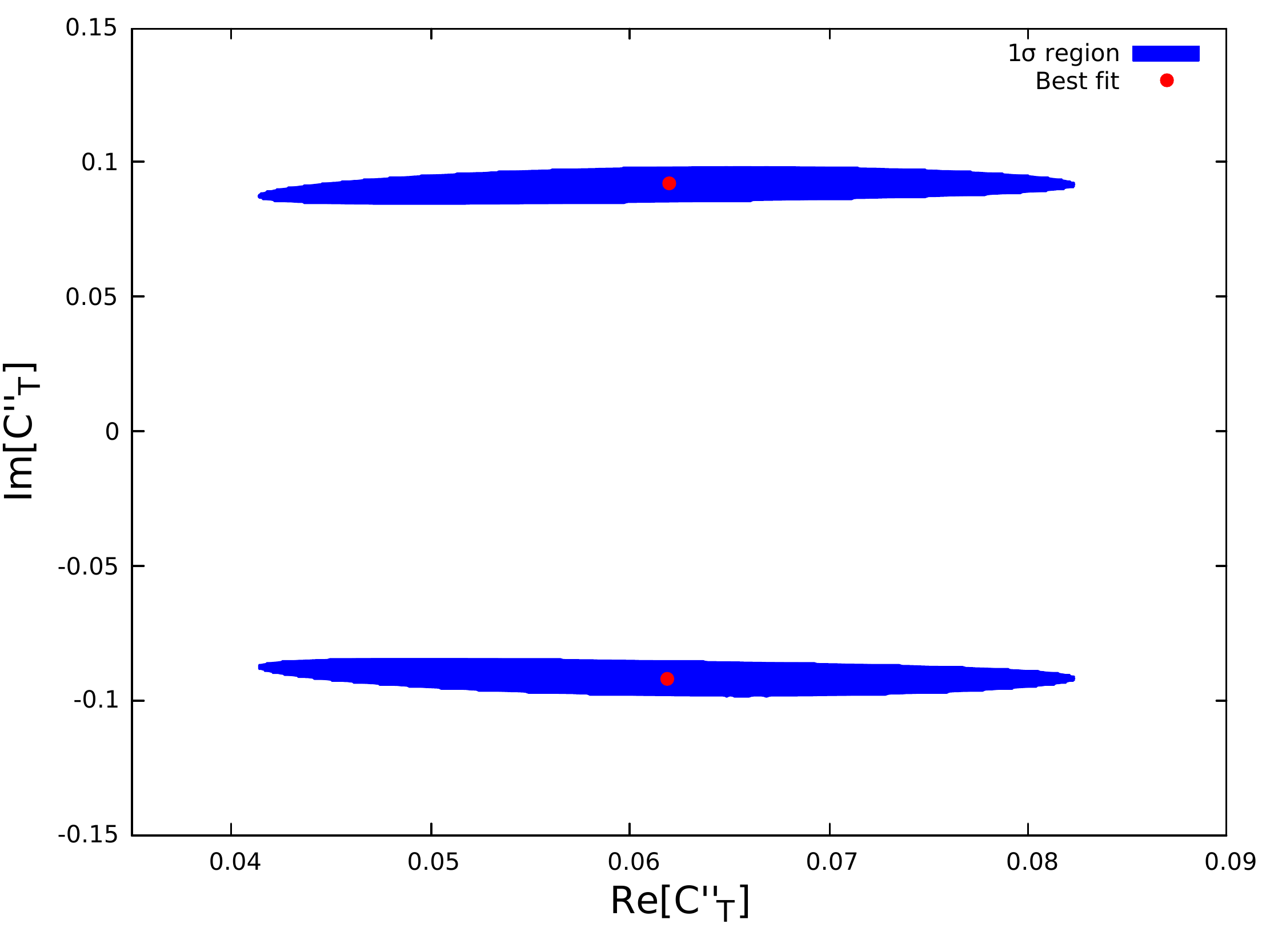}&
\includegraphics[width=75mm]{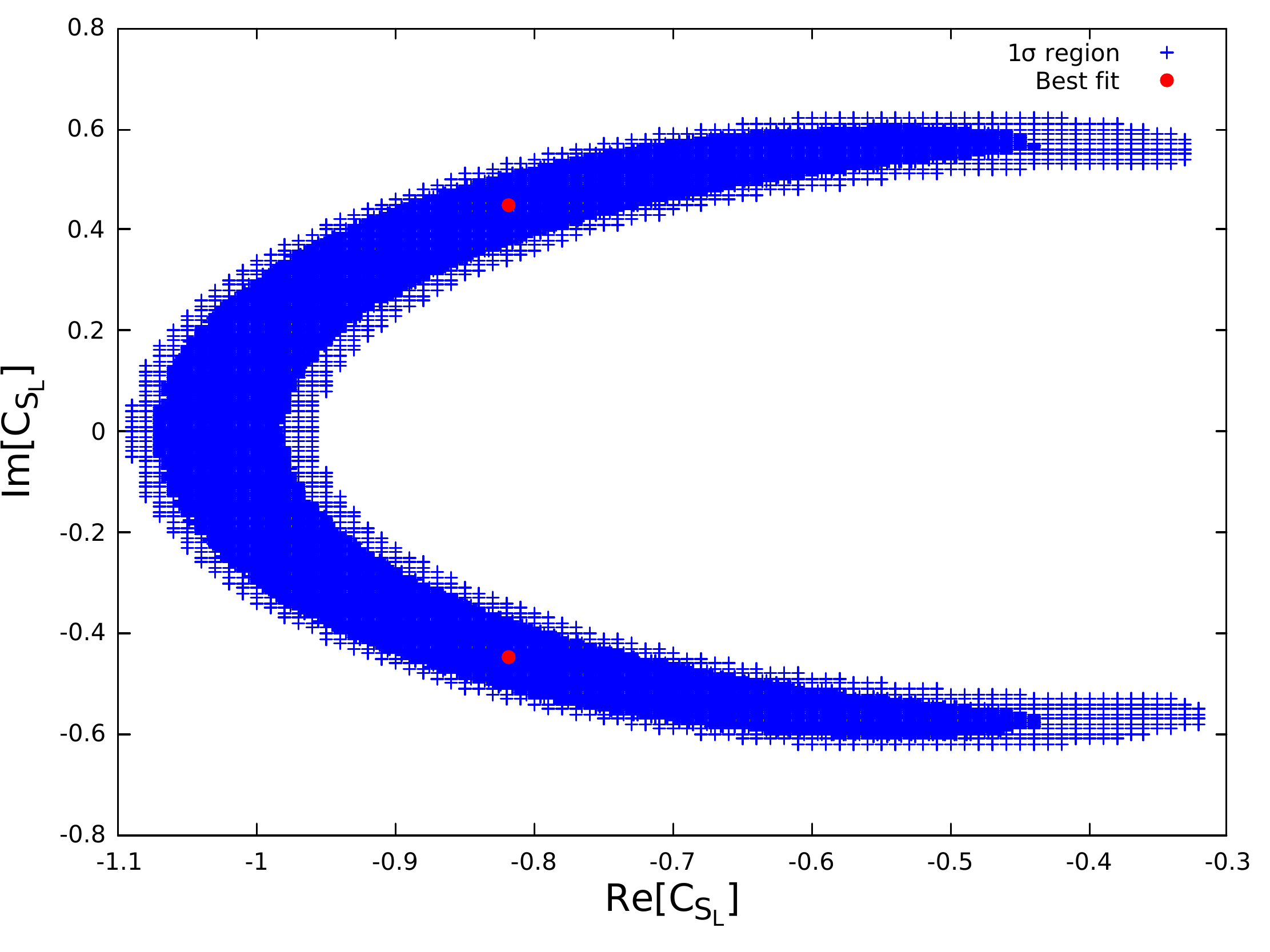}\\
\end{tabular}
\caption{The allowed $1\sigma$ regions for the complex NP WCs listed in Table~\ref{tab1}. For each plot, the blue colored region corresponds to the $1\sigma$ parameter space whereas the red dots represents the best fit values of NP WCs.}
\label{fig1}
\end{figure*}

To obtain the values of NP WCs, we minimize the $\chi^2$ function by taking non-zero value of one NP WC at a time. While doing so, we set other coefficients to be zero. This minimizations is performed by the CERN $\tt MINUIT$ library \cite{James:1975dr,James:1994vla}. We find that the values of $\chi^2_{\rm min}$ fall into two disjoint ranges $\lesssim 4.5$ and $\gtrsim 9$. We keep only those NP WCs which satisfy $\chi^2_{\rm min}\leq 4.5$. The central values of these allowed WCs of NP solutions are listed in Table \ref{tab1}. We do not provide the errors of individual best fit values because of the correlation between the real and imaginary parts. In stead, we show the $1\sigma$ allowed regions for theses NP solutions in Fig.~\ref{fig1}. 
\begin{table}[htbp]
\centering
\tabcolsep 6pt

\begin{tabular}{|c|c|c|c|}
\hline\hline
NP type &  Best fit value(s)  & $\chi^2_{\rm min}$ & pull \\
\hline
$C_{V_L}$ & $0.10\pm 0.12\,i$& $4.55$ & $4.1$\\
\hline
$C'_{S_L}$ & $0.25\pm 0.86\,i$ & $4.50$ & $4.2$\\
\hline
$C''_T$ & $0.06\pm 0.09\,i$ & $3.45$ & $4.3$\\
\hline
$C_{S_L}$ & $-0.82\pm 0.45\,i$ & $2.50$ & $4.4$\\

\hline\hline
\end{tabular}
\caption{Best fit values of NP WCs at $\Lambda = 1$ TeV for the measurements of  $R_D$, $R_{D^*}$, $R_{J/\psi}$, $P^{D^*}_{\tau}$ and $F^{D^*}_L$. We list the central values of the NP solutions with $\chi^2_{\rm min}\leq 4.5$. For the SM, we have $\chi^2_{\rm SM} = 21.80$. The pull values are calculated using pull = $\sqrt{\chi^2_{\rm SM}-\chi^2_{\rm min}}$.}
\label{tab1}
\end{table}

The purely leptonic decay $B_c\rightarrow \tau\bar{\nu}$ plays a crucial role to constrain the NP solutions in this sector. This decay is subject to helicity suppression in the SM whereas this suppression is removed for the pseudo-scalar operators. Therefore, these NP operators are highly constrained by this observable. 
Within the NP framework, the branching fraction of $B_c\rightarrow \tau\bar{\nu}$ can be expressed as 
\begin{eqnarray}
Br(B_c\rightarrow \tau\bar{\nu}) &=& \frac{\vert V_{cb}\vert^2G^2_Ff^2_{B_c}m_{B_c}m^2_{\tau}\tau^{exp}_{B_c}}{8\pi}\left(1-\frac{m^2_{\tau}}{m^2_{B_c}}\right)^2\times \nonumber\\
& &  \left| 1+C_{V_L}-C_{V_R}+\frac{m^2_{B_c}}{m_{\tau}(m_b+m_c)}(C_{S_R}-C_{S_L})\right|^2,
\end{eqnarray}
where the decay constant $f_{B_c}= 434\pm 15$ MeV \cite{Colquhoun:2015oha} and the measured lifetime  $\tau^{exp}_{B_c}=0.510\pm 0.009$ ps \cite{Tanabashi:2018oca}. Here  $m_b$ and $m_c$ are the running quark masses evaluated at the $\mu_b=m_b$ scale. The SM predicts this branching fraction to be $\sim 2.15\times 10^{-2}$. 

In Ref.~\cite{Akeroyd:2017mhr}, the upper limit on this branching ratio is set to be $10\%$ from the LEP data which are admixture of $B_c\to \tau\bar{\nu}$ and $B_u\to \tau\bar{\nu}$ decays at $Z$ peak. To extract the $Br(B_c\to \tau\bar{\nu})$, one needs to know the ratio of fragmentation functions of $B_c$ and $B_u$ mesons defined as $f_c/f_u$. The value of this ratio is obtained from the data of Tevatron~\cite{Abe:1998fb,Abulencia:2006zu} and LHCb~\cite{Aaij:2014jxa}. On the other hand, the authors of Ref.~\cite{Alonso:2016oyd} obtained this upper limit to be $30\%$ by making use of the lifetime of $B_c$ meson. This is estimated by considering that the $B_c\to \tau \bar{\nu}$ decay rate does not exceed the fraction of the total width which is allowed by the calculation of the lifetime in the SM. In Ref.~\cite{Blanke:2019qrx}, the authors have argued that these two different upper limits are too conservative and these could be over-estimated. However, taking all uncertainties into account the decay width of $B_c$ meson can be relaxed up to $60\%$ which is not that much conservative. Therefore, we consider these three different upper limits on branching ratio of $B_c\rightarrow \tau\bar{\nu}$ to constrain the NP parameter space. In this analysis, the NP WCs are defined at a scale $\Lambda$= 1 TeV. However, all these physical processes happen at $m_b$ scale.  Therefore, we include the renormalization group (RG) effects in the evolution of the WCs from the scale of $1$ TeV to the $m_b$ scale~\cite{Gonzalez-Alonso:2017iyc}. In particular, these effects are important for the scalar and tensor operators. 

\begin{figure*}[htbp] 
\centering
\begin{tabular}{cc}
\includegraphics[width=75mm]{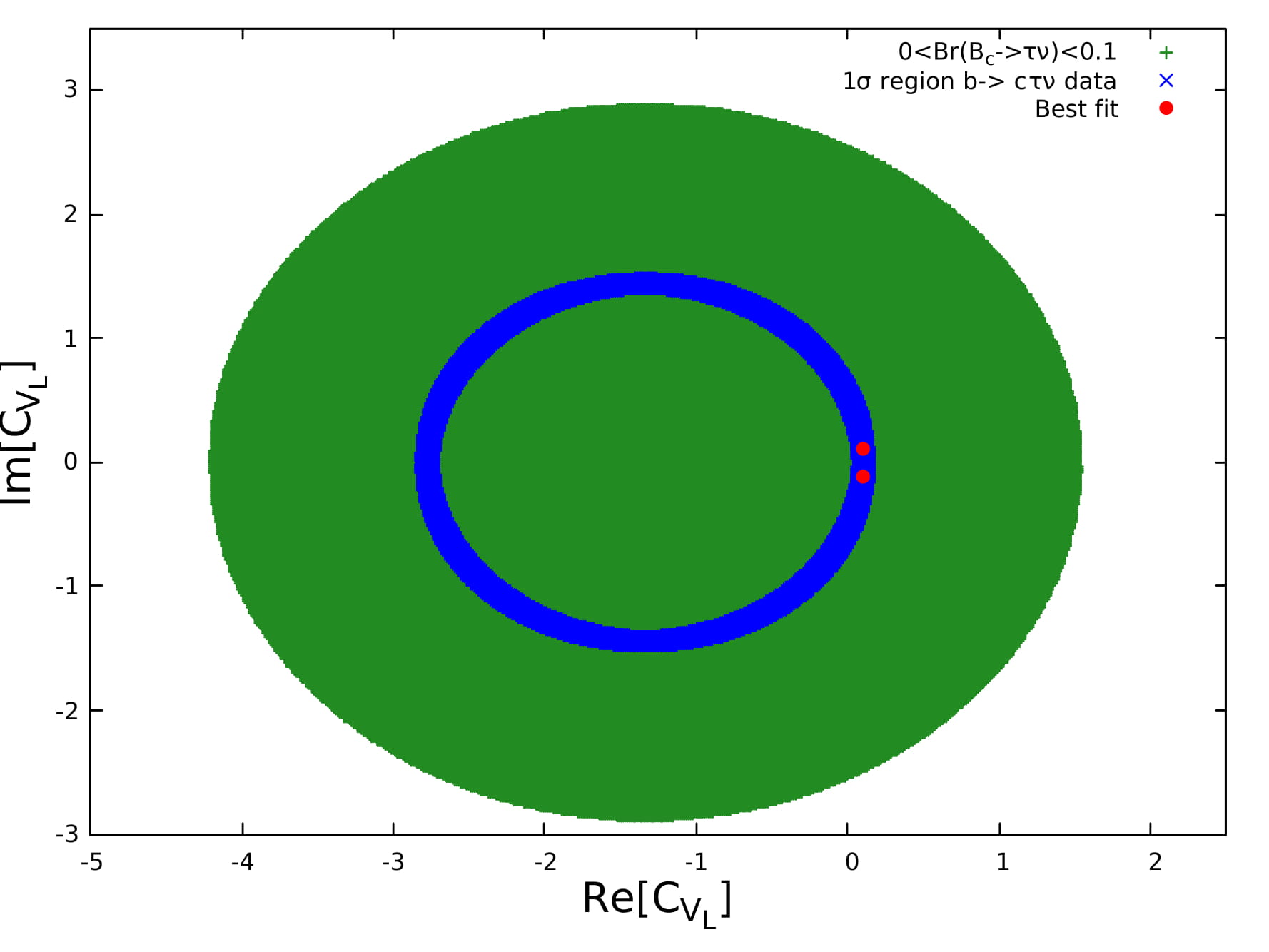}&
\includegraphics[width=75mm]{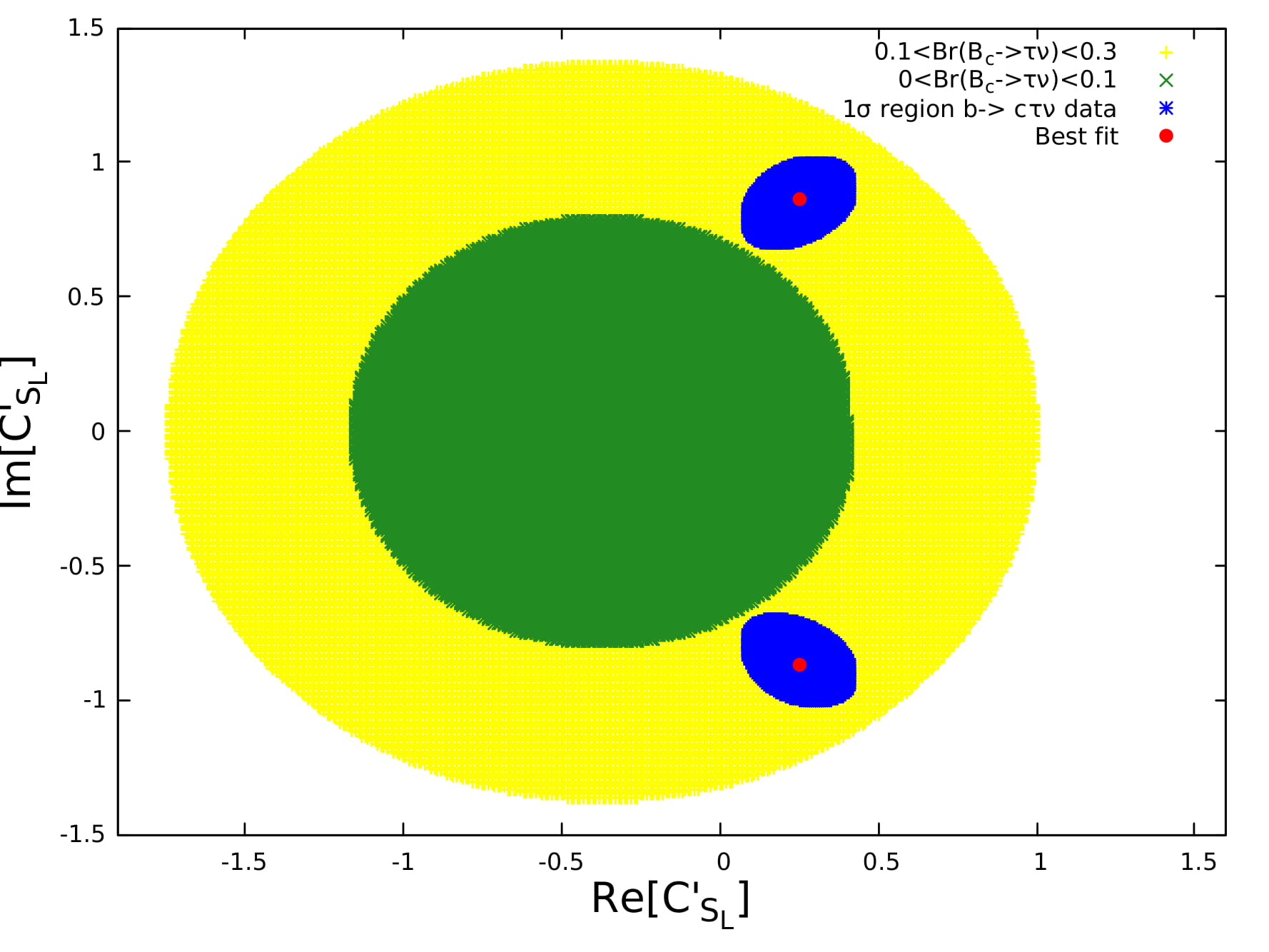}\\
\includegraphics[width=77mm]{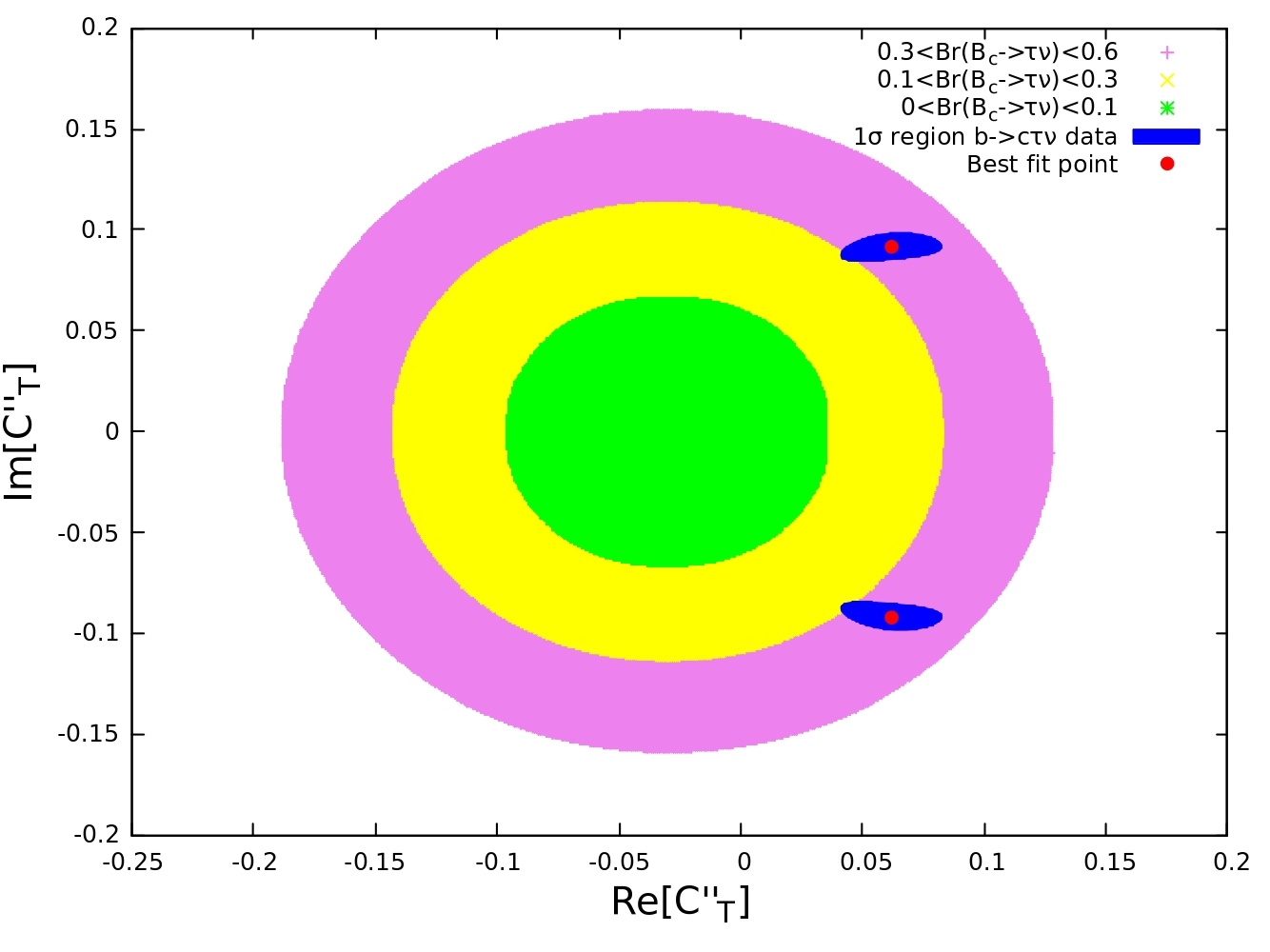}&
\includegraphics[width=77mm]{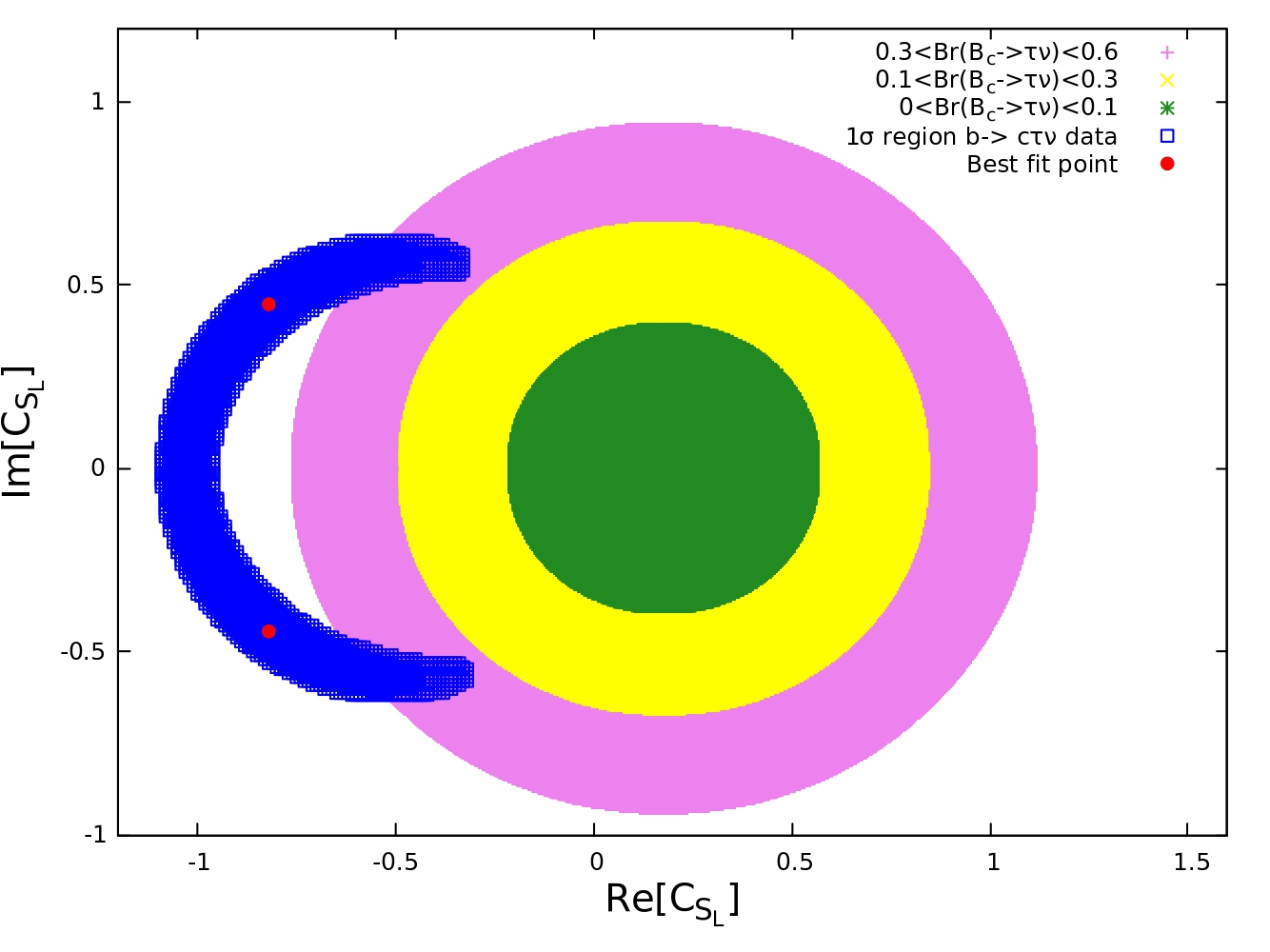}\\
\end{tabular}
\caption{The $1\sigma$ regions allowed by $b\to c\tau\bar{\nu}$ data (blue) and parameter spaces for three different upper limits $10\%$ (green), $30\%$ (yellow), $60\%$ (violet) of $Br(B_c\rightarrow \tau\bar{\nu})$ for each complex NP WC listed in Table~\ref{tab1}. In each plot, the red dots represent the best fit points. }
\label{fig2}
\end{figure*}

In Fig.~\ref{fig2}, we have shown the parameter space which span $1\sigma$ region allowed by present $b\rightarrow c\tau\bar{\nu}$ data and by the three different upper limits on the branching ratio of $B_c\rightarrow \tau\bar{\nu}$. The best fit point for each solution listed in Table~\ref{tab1} is also plotted within the $1\sigma$ allowed region. Only the $\mathcal{O}_{V_L}$ solution falls within the allowed space constrained by $Br(B_c\rightarrow \tau\bar{\nu})<10\%$. The allowed $1\sigma$ regions for $\mathcal{O}'_{S_L}$ and $\mathcal{O}''_T$ solutions fall into the regions allowed by the constraints $Br(B_c\rightarrow \tau\bar{\nu})$ $< 30\%$ and $< 60\%$ respectively. The best fit NP WCs of $\mathcal{O}_{S_L}$ solution do not fall into the region allowed by the constraint  $Br(B_c\rightarrow \tau\bar{\nu})<60\%$ whereas a small fraction of the $1\sigma$ region overlaps with the region allowed by $30\% <Br(B_c\rightarrow \tau\bar{\nu})<60\%$. Hence we can reject the mildly allowed $\mathcal{O}_{S_L}$ solution. We list the final three allowed NP solutions in Table~\ref{tab2}. 

Using the best fit values of the allowed solutions, we provide the predicted central values of the quantities used in the fit, i.e., $R_D, R_{D^*}, R_{J/\psi}$, $P^{D^*}_{\tau}$ and $F_L^{D^*}$, for each solution. This will allow us to see how close are the predictions of NP solutions to the experimental measurements.
We note the following observations by looking at the predictions in Table~\ref{tab2}:
\begin{itemize}
\item The predictions of $R_D$, $R_{D^*}$ and $P^{D^*}_{\tau}$ for the three solutions are within 1$\sigma$ of the respective experimental averages.
\item The predicted values of $R_{J/\psi}$ and $F^{D^*}_L$ for the three solutions are within $\sim 1.6\sigma$ of the experimental measurements. The Lorentz structure of $\mathcal{O}'_{S_L}$ is different from that of $\mathcal{O}_{V_L}$. But the prediction of $F^{D^*}_L$ for $\mathcal{O}'_{S_L}$ solution is the same as that of $\mathcal{O}_{V_L}$ solution because the value of WC is quite small. However, these two NP solutions fall in two different ranges of $Br(B_c\to \tau\bar{\nu})$ because the helicity suppression is lifted in presence of the $\mathcal{O}'_{S_L}$ solution. 
\end{itemize}
\begin{table}[htbp]
\centering
\tabcolsep 2pt
\begin{tabular}{|c|c|c|c|c|c|c|}
\hline\hline
NP type & Best fit value(s)  & $R_D$ & $R_{D^*}$ & $R_{J/\psi}$ & $P^{D^*}_{\tau}$ & $F^{D^*}_L$\\
\hline
SM  & $C_{i}=0$  &$0.297\pm 0.008$ &$0.253\pm 0.002$ & $0.289\pm 0.008$ &$-0.499\pm 0.004$ & $0.457\pm 0.005$ \\
\hline
$C_{V_L}|_{10\%}$  &  $0.10\pm 0.12\, i$  &$0.364\pm 0.010$ &$0.294\pm 0.005$ & $0.334\pm 0.010$ & $-0.499\pm 0.005$ & $0.443\pm 0.007$ \\
\hline
$C'_{S_L}|_{30\%}$ & $0.25 \pm 0.86\, i$ & $0.336\pm 0.009$ & $0.295\pm 0.005$ & $0.339\pm 0.011$ & $-0.419\pm 0.006$ & $0.443\pm 0.007$\\
\hline
$C''_T|_{60\%}$ & $0.06 \pm 0.09\,i$ &$0.333\pm 0.010$ & $0.296\pm 0.006$ & $0.344\pm 0.009$ & $-0.375\pm 0.005$ & $0.420\pm 0.006$\\ 
\hline\hline
\end{tabular}
\caption{Central values of best fit NP WCs at $\Lambda = 1$ TeV by making use of data of $R_D$, $R_{D^*}$, $R_{J/\psi}$, $P^{D^*}_{\tau}$ and $F^{D^*}_L$. Here we allow only those solutions for which $\chi^2_{\rm min}\leq 4.5$ as well as for three different upper limits $10\%$, $30\%$ and $60\%$ of $Br(B_c\rightarrow \tau\bar{\nu})$. We also provide the predictions of each observables which are taken into the fit.}
\label{tab2}
\end{table}
\begin{table}[ht]
\centering
\tabcolsep 6pt
\begin{tabular}{|c|c|c|c|c|}
\hline\hline
NP type & Best fit value(s)  & $P^D_{\tau}$ & $A^D_{FB}$ & $A^{D^*}_{FB}$ \\
\hline
SM  & $C_{i}=0$  & $0.325\pm 0.001$ & $0.360\pm 0.002$  & $-0.063\pm 0.005$ \\
\hline
$C_{V_L}|_{10\%}$  &  $0.10\pm 0.12\, i$  & $0.325\pm 0.002$ & $0.360\pm 0.002$  & $-0.063\pm 0.006$ \\
\hline
$C'_{S_L}|_{30\%}$ & $0.25 \pm 0.86\, i$ & $0.420\pm 0.001$ & $0.212\pm 0.003$ & $0.0001\pm 0.005$\\
\hline
$C''_T|_{60\%}$ & $0.06 \pm 0.09\,i$ & $0.414\pm 0.002$ & $0.100\pm 0.004$ & $0.009\pm 0.006$\\ 
\hline\hline
\end{tabular}
\caption{Average values of angular observables $P^D_{\tau}$, $A^D_{FB}$ and $A^{D^*}_{FB}$ for the SM and three solutions listed in Table~\ref{tab2}.}
\label{tab3}
\end{table}

We consider other angular observables in $B\to (D,D^*)\tau\bar{\nu}$ decay which are yet to be measured. In particular, we are interested in the following three observables~\cite{Alok:2018uft}
\begin{itemize}
\item The polarization of $\tau$ lepton in $B\to D\tau\bar{\nu}$ decay, $P^D_{\tau}$
\item The forward-backward asymmetry in $B\to D\tau\bar{\nu}$ decay, $A^D_{FB}$ and 
\item The forward-backward asymmetry in $B\to D^*\tau\bar{\nu}$ decay, $A^{D^*}_{FB}$.
\end{itemize} 
\begin{figure*}[ht] 
\centering
\begin{tabular}{cc}
\includegraphics[width=75mm]{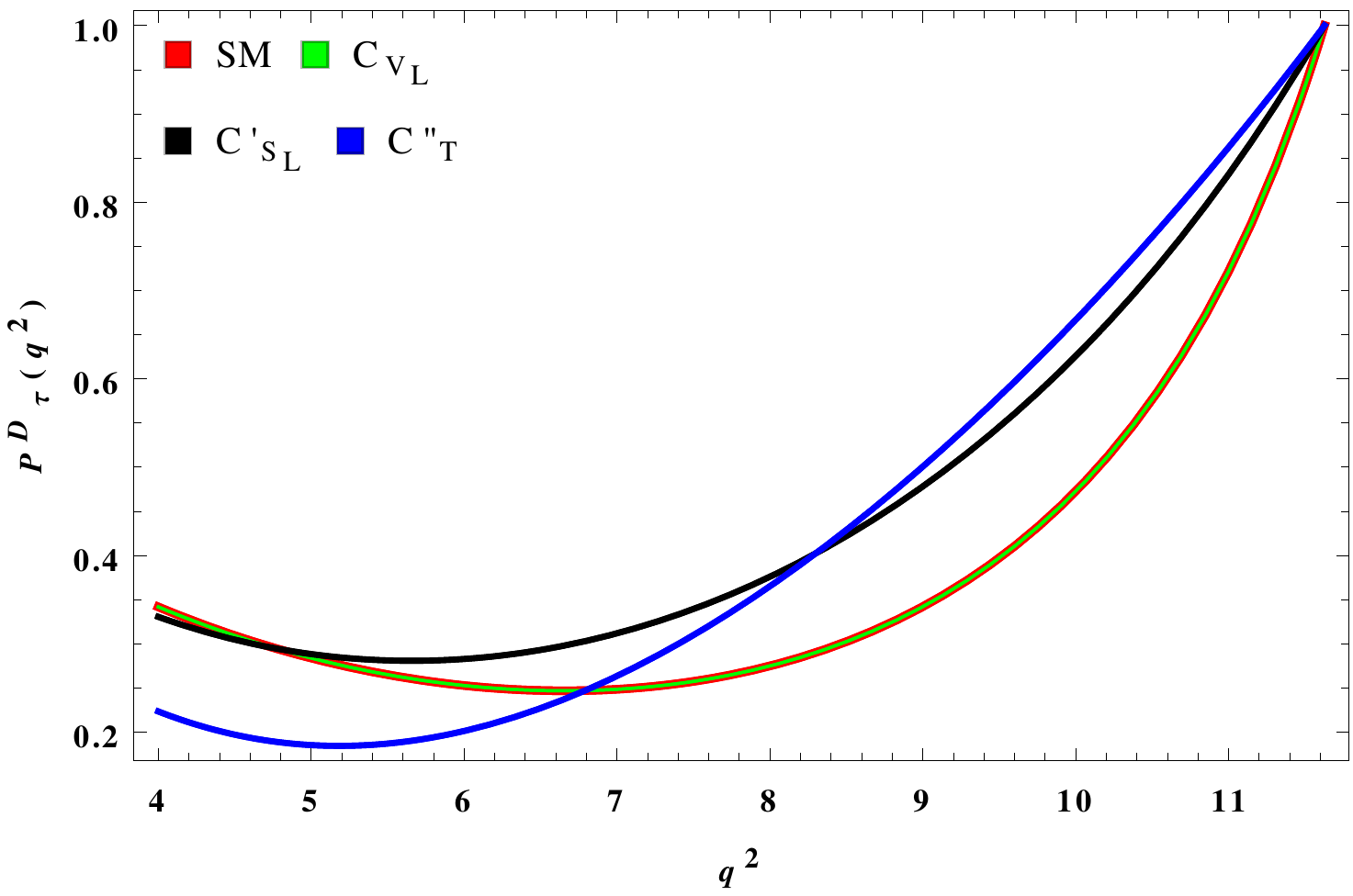}&
\includegraphics[width=75mm]{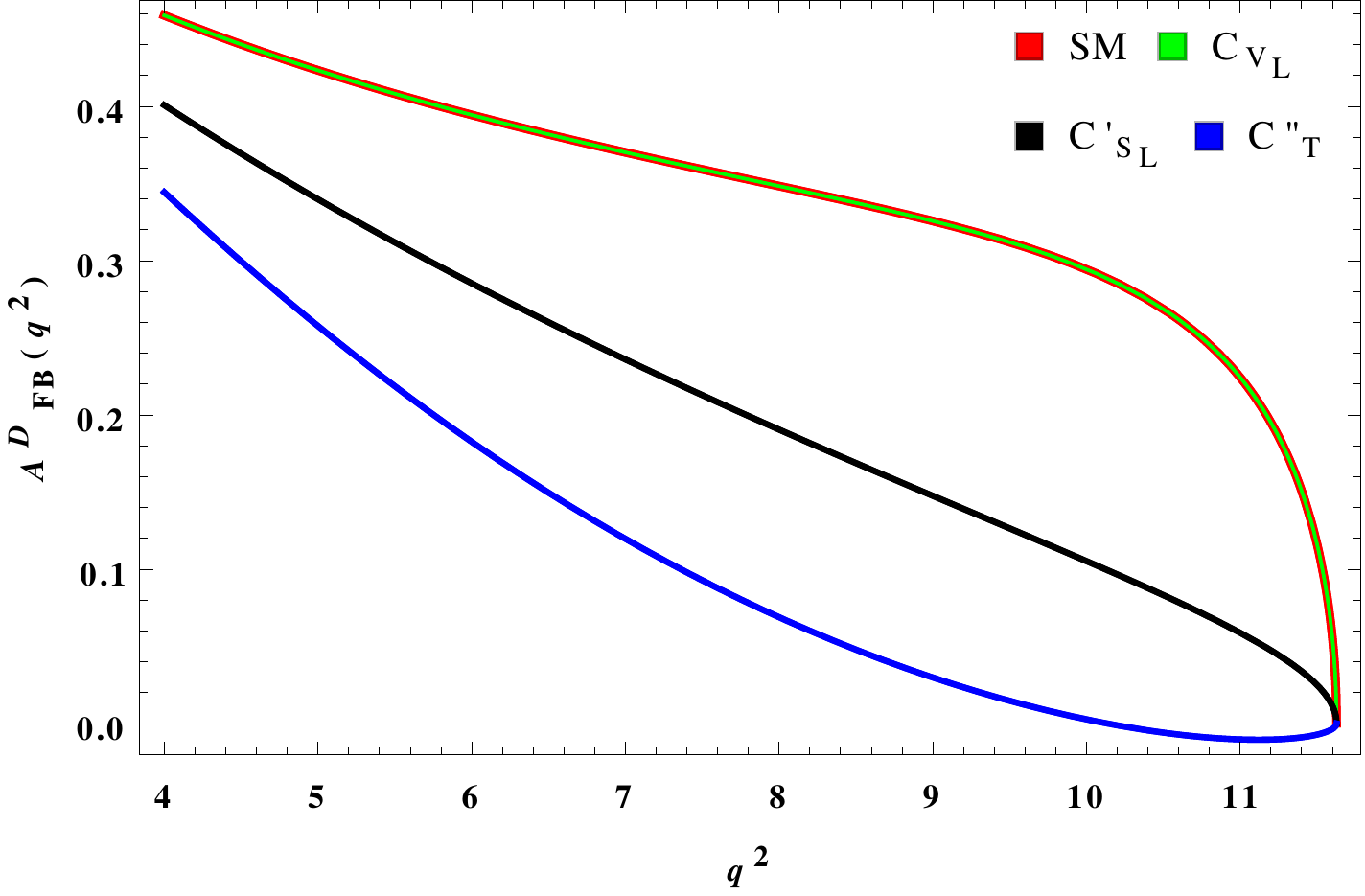}\\
\end{tabular}
\includegraphics[width=75mm]{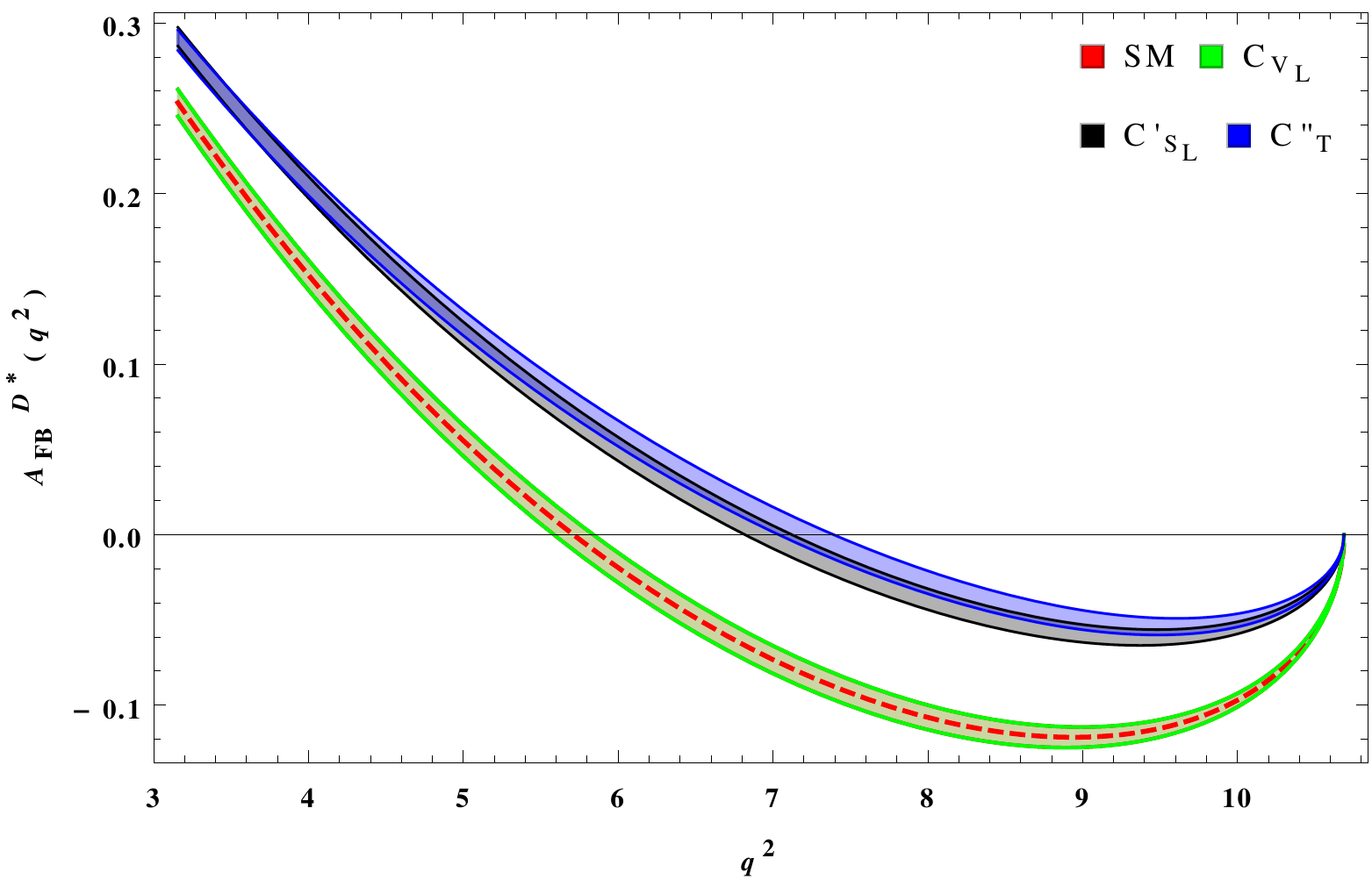}\\
\caption{The predictions of angular observables $P^D_{\tau}$, $A^D_{FB}$ and $A^{D^*}_{FB}$ as a function of $q^2$ (GeV$^2$) for the SM and three solutions listed in Table~\ref{tab3}. The color code for each case is shown in each plot.}
\label{fig3}
\end{figure*}
We compute the average values of these three angular observables for the allowed NP solutions. The predicted values are listed in Table~\ref{tab3}. For completeness, we also plot these observables as a function of $q^2=(p_B-p_{D^{(*)}})^2$, where $p_B$ and $p_{D^{(*)}}$ are the respective four momenta of $B$ and $D^{(*)}$ mesons. These are shown in Fig.~\ref{fig3}. From Table~\ref{tab3} and Fig.~\ref{fig3}, we observe the following features
\begin{itemize}
\item The predictions of all three observables for the $\mathcal{O}_{V_L}$ solution are exactly same as those of the SM. This is because the Lorentz structure of $\mathcal{O}_{V_L}$ operator is same as the SM.
\item The $P^D_{\tau}$ has very poor discriminating capability.
\item The predictions of $A^D_{FB}$ and $A^{D^*}_{FB}$ for the $\mathcal{O}'_{S_L}$ and $\mathcal{O}''_{T}$ solutions are markedly different. These two solutions can be distinguished by forward-backward asymmetries.
\end{itemize}


\section{CP violating triple product asymmetries}
If the hints of LFU violation in $b\to c\tau\bar{\nu}$ sector is indeed due to new physics, then it is very likely that the new
physics will contain additional phases which can lead to some signatures of CP violation in the relevant decay modes. In this section, we discuss about the possible CP violation in $B\to D^*\tau\bar{\nu}$ decay. The simplest possible CP violating observable, which one could think of, is the direct CP asymmetry between the decay and its CP conjugate mode. In order to have a non-zero value of direct CP asymmetry, we need strong phase difference between the amplitudes besides the weak phase. For $B\to D^*\tau\bar{\nu}$ decay, there is no strong phase difference in the SM because of unique final state of the decay and its CP conjugate mode. In Ref.~\cite{Aloni:2018ipm}, the authors suggested a mechanism where this strong phase difference could arise due to interference between the higher resonances of $D^*$ meson. They have shown that the CP violation could be as large as $\sim 10\%$ only for the tensor NP. However, the tensor NP is now ruled out by the Belle measurement on $F^{D^*}_L$.

In this work, we focus on CP violating triple product asymmetries (TPA) in $B\to D^*\tau\bar{\nu}$ decay. The full angular distribution of quasi-four body decay $B\to D^*(\to D\pi)\tau\bar{\nu}$ can be described by four independent parameters $-$ (a) $q^2 = (p_B-p_{D^*})^2$ where $p_B$ and $p_{D^*}$ are respective four momenta of $B$ and $D^*$ meson, (b) $\theta_{D}$ the angle between $B$ and $D$ mesons where $D$ meson
comes from $D^*$ decay, (c) $\theta_{\tau}$ the angle between $\tau$ momenta and $B$ meson, and (d) $\phi$ the angle between $D^*$ decay plane and the plane defined by the $\tau$ and $\nu$ momenta~\cite{Duraisamy:2014sna}. The triple products (TP) are obtained by integrating the full decay distribution in different ranges of the polar angles $\theta_D$ and $\theta_{\tau}$. These are following~\cite{Alok:2011gv,Duraisamy:2013kcw,Bhattacharya:2019olg,Bhattacharya:2020lfm}
\begin{eqnarray}
\frac{d^2\Gamma^{(1)}}{dq^2d\phi} &= & \int^1_{-1} \int^1_{-1} \frac{d^4\Gamma}{dq^2 d\cos\theta_{\tau}d\cos\theta_{D}d\phi} d\cos\theta_{\tau}d\cos\theta_{D} \\ \nonumber
&= &\frac{1}{2\pi}\frac{d\Gamma}{dq^2}\left[1+\left(A^{(1)}_C\,\cos 2\phi + A^{(1)}_T\,\sin 2\phi\right)\right],
\end{eqnarray}
\begin{eqnarray}
\frac{d^2\Gamma^{(2)}}{dq^2d\phi} &=& \int^1_{-1} d\cos\theta_{\tau} \left[\int^1_0 -\int^0_{-1} \right] \frac{d^4\Gamma}{dq^2 d\cos\theta_{\tau}d\cos\theta_{D}d\phi} d\cos\theta_{D} \nonumber \\
&=& \frac{1}{4}\frac{d\Gamma}{dq^2}\left[A^{(2)}_C\,\cos \phi + A^{(2)}_T\,\sin \phi\right],
\end{eqnarray}
and
\begin{eqnarray}
\frac{d^2\Gamma^{(3)}}{dq^2d\phi} &=&\left[\int^1_0 -\int^0_{-1} \right] d\cos\theta_{\tau}  \left[\int^1_0 -\int^0_{-1} \right] \frac{d^4\Gamma}{dq^2 d\cos\theta_{\tau}d\cos\theta_{D}d\phi} d\cos\theta_{D} \nonumber \\
&= & \frac{2}{3\pi}\frac{d\Gamma}{dq^2}\left[A^{(3)}_C\,\cos \phi + A^{(3)}_T\,\sin \phi\right].
\end{eqnarray}
The coefficients $A^{(i)}_C$ of $\cos\phi$ and $\cos 2\phi$ are even under CP transformation and hence we are not interested in these. However, the angular coefficients $A^{(i)}_T$ of $\sin \phi$ and $\sin 2\phi$ are odd under the CP transformation which leads to these quantities to be CP violating observables. These three TPs are defined as follows~\cite{Duraisamy:2013kcw}:
\begin{equation}
A^{(1)}_T(q^2) = \frac{4V^T_5}{A_L+A_T}, \quad A^{(2)}_T(q^2) = \frac{V^{0T}_3}{A_L+A_T}, \quad	A^{(3)}_T(q^2) = \frac{V^{0T}_4}{A_L+A_T},
\label{TPs}
\end{equation}
where $V$'s are the angular coefficients and $A_L$ and $A_T$ are the longitudinal and transverse amplitudes respectively. The expressions for these quantities are given in Appendix~\ref{app} and also can be found in Ref.~\cite{Duraisamy:2014sna}. The SM predictions of these TPs are almost zero. Therefore, the complex NP WCs can predict a non-zero value for these quantities. Thus these TPs provide a new degree of freedom to test beyond SM physics. For the CP conjugate decay, the definitions in Eq.~(\ref{TPs}) take the following forms
\begin{equation}
\bar{A}^{(1)}_T(q^2) = -\frac{4\bar{V}^T_5}{\bar{A}_L+\bar{A}_T}, \quad \bar{A}^{(2)}_T(q^2) = \frac{\bar{V}^{0T}_3}{\bar{A}_L+\bar{A}_T}, \quad	\bar{A}^{(3)}_T(q^2) = -\frac{\bar{V}^{0T}_4}{\bar{A}_L+\bar{A}_T}.
\label{TPsbar}
\end{equation}
Using Eqs.~(\ref{TPs}) and (\ref{TPsbar}), three asymmetries can be defined between the corresponding TPs of the decay and its CP conjugate. These TPAs are defined as follows
\begin{eqnarray}
\langle A^{(1)}_T(q^2)\rangle &=& \frac{1}{2}\left(A^{(1)}_T(q^2) + \bar{A}^{(1)}_T(q^2)\right), \nonumber \\
\langle A^{(2)}_T(q^2)\rangle &=& \frac{1}{2}\left(A^{(2)}_T(q^2) - \bar{A}^{(2)}_T(q^2)\right), \nonumber \\
\langle A^{(3)}_T(q^2)\rangle &=& \frac{1}{2}\left(A^{(3)}_T(q^2) + \bar{A}^{(3)}_T(q^2)\right).
\end{eqnarray}
\begin{figure*}[ht] 
\centering
\begin{tabular}{cc}
\includegraphics[width=75mm]{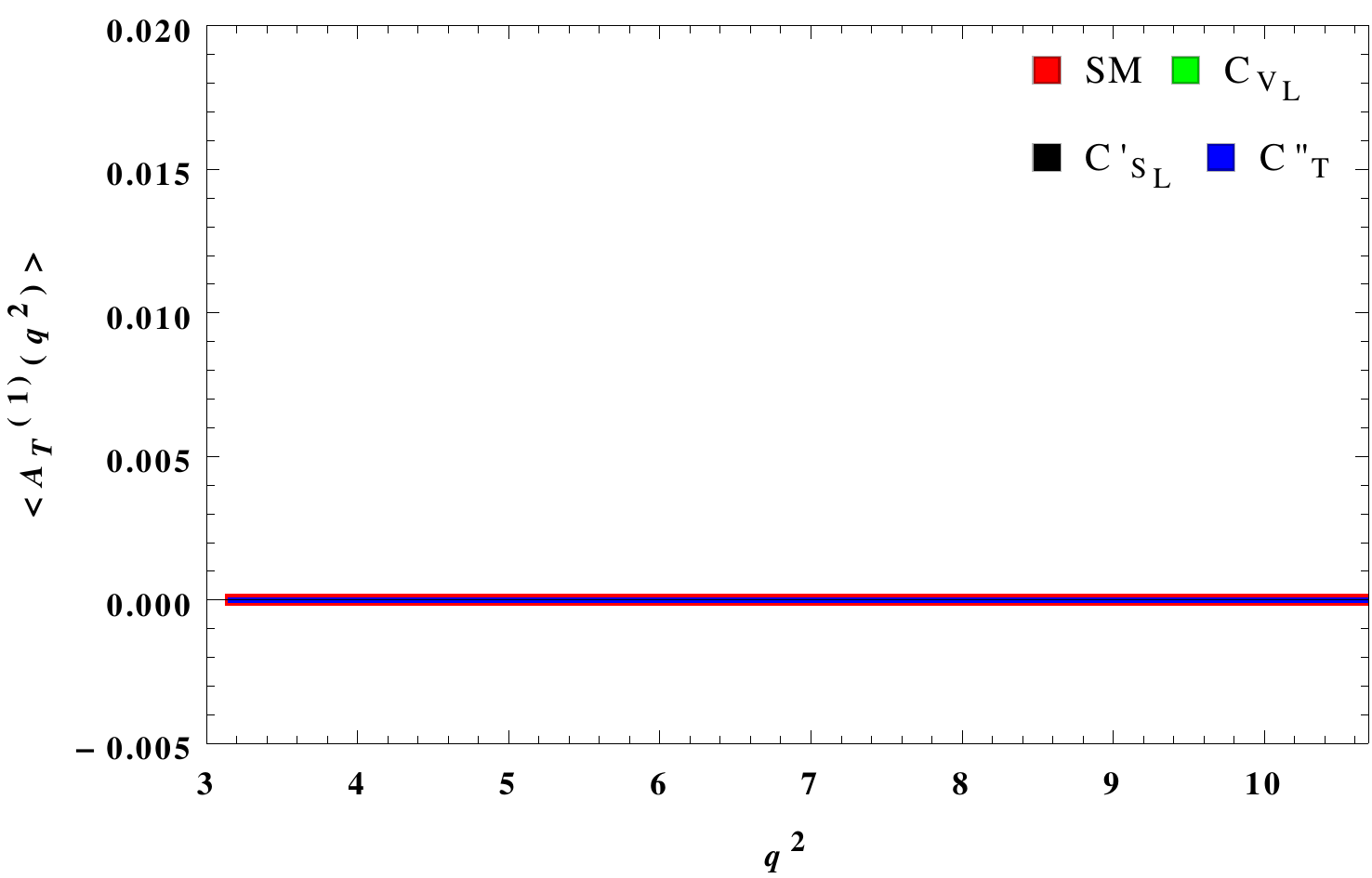}&
\includegraphics[width=75mm]{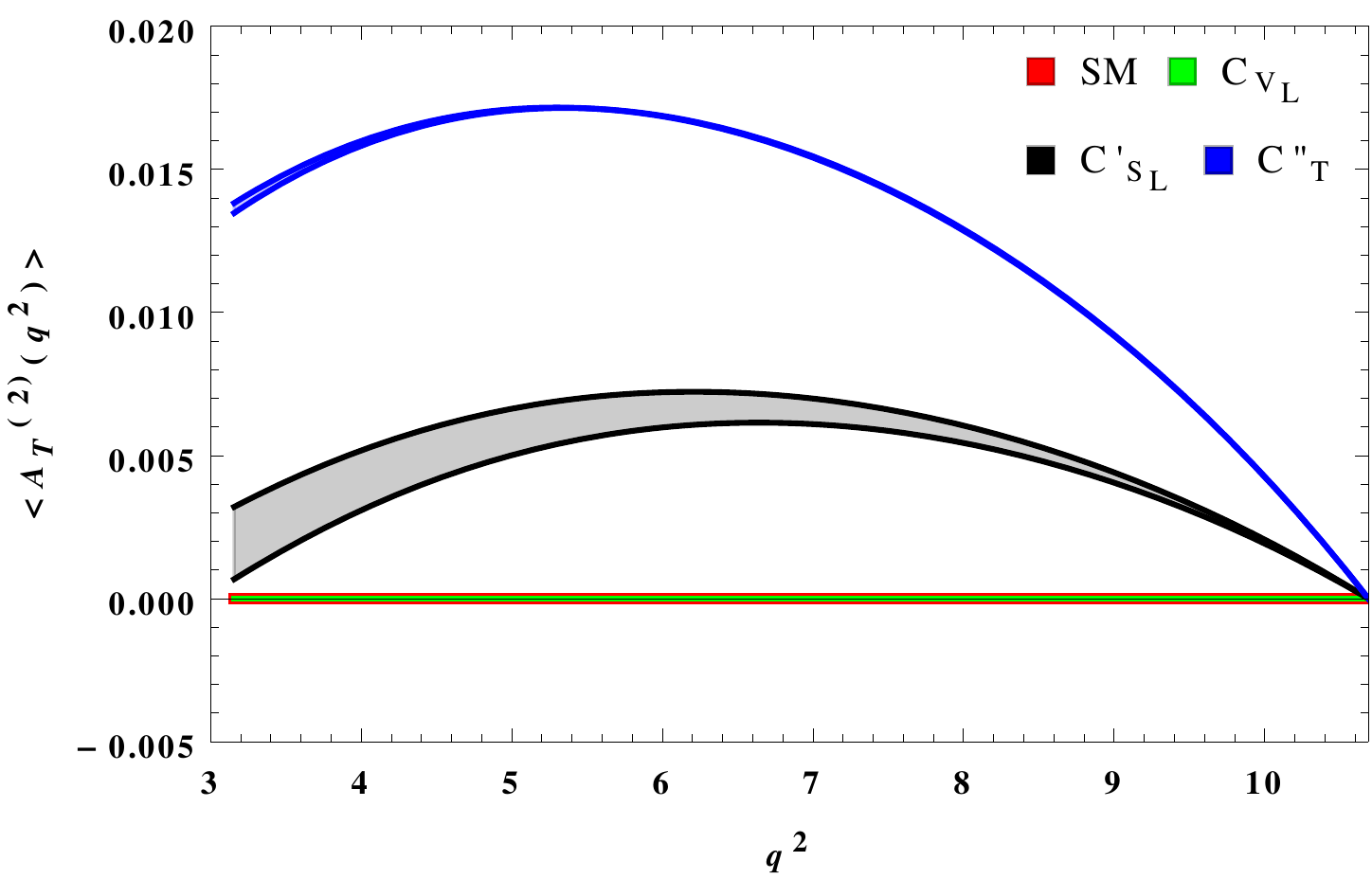}\\
\end{tabular}
\includegraphics[width=75mm]{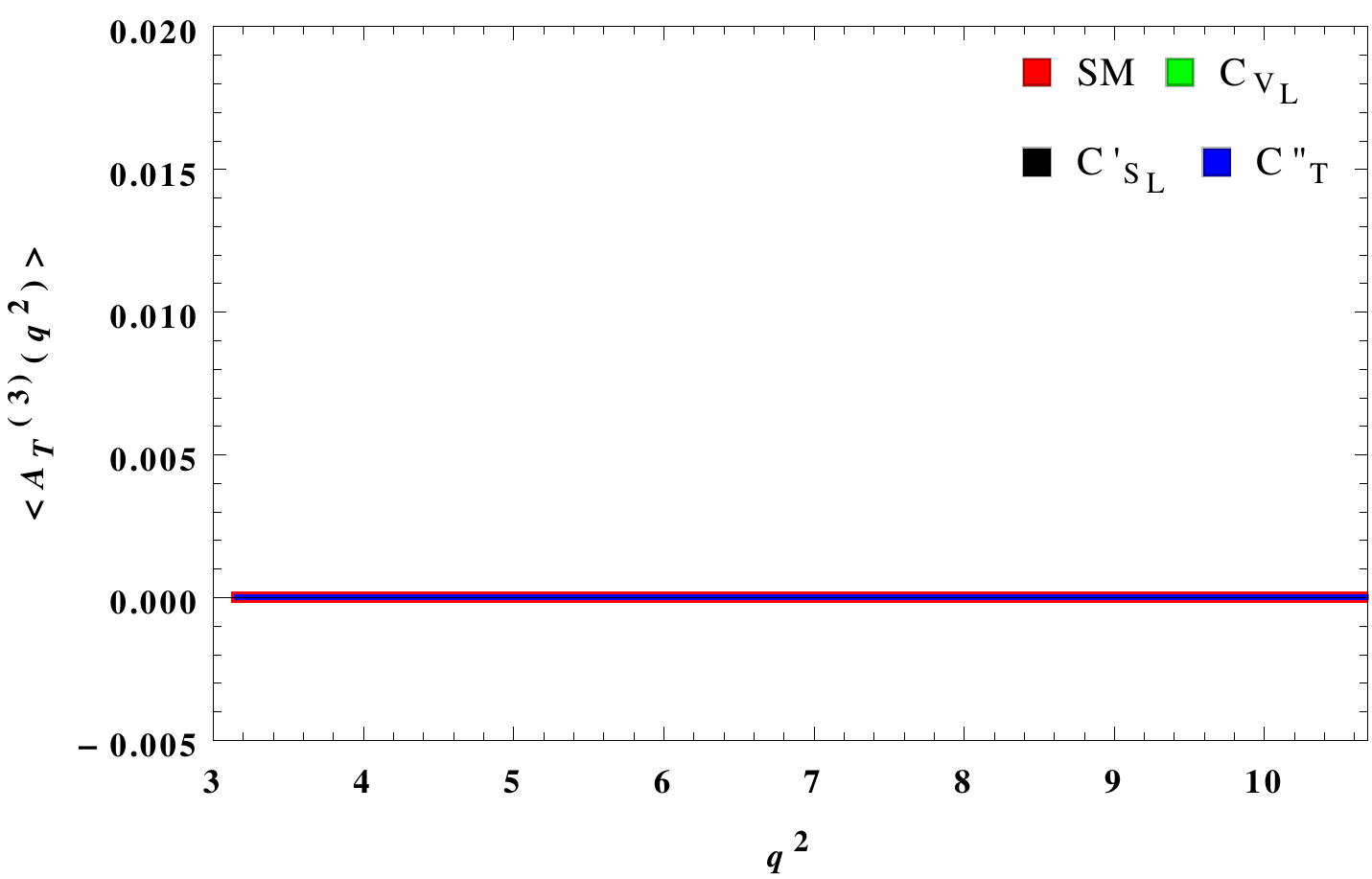}\\
\caption{The TPAs are plotted as a function of $q^2$ (GeV$^2$) for the SM and three best fit NP WCs listed in Table~\ref{tab2}. The color code for each plot is shown in figure.}
\label{fig4}
\end{figure*}
First we calculate the predictions of these TPAs for the SM and the three best fit NP solutions listed in Table~\ref{tab2} as a function of $q^2$. These predictions are shown in Fig.~\ref{fig4}. From this figure, we make the following observations
\begin{itemize}
\item The TPAs $\langle A^{(1)}_T(q^2)\rangle$ and $\langle A^{(3)}_T(q^2)\rangle$ depend only on the $\mathcal{O}_{V_L}$ and $\mathcal{O}_{V_R}$ operators. The $\mathcal{O}_{V_L}$ has the same Lorentz structure as the SM. Therefore, the $\mathcal{O}_{V_L}$ solution predicts these two asymmetries to be zero for whole $q^2$ range. For other two NP solutions, the predictions are zero because these two asymmetries do not depend on those NP WCs.  
\item The TPA $\langle A^{(2)}_T(q^2)\rangle$ depends on $\mathcal{O}_{V_L}$, $\mathcal{O}_{V_R}$, $\mathcal{O}_{S_L}$, $\mathcal{O}_{S_R}$ and $\mathcal{O}_T$ operators. The $\mathcal{O}_{V_L}$ operator has the same Lorentz structure as the SM. Hence, the prediction of this TPA is zero for  the $\mathcal{O}_{V_L}$ solution for whole $q^2$ range. The $\mathcal{O}'_{S_L}$ and $\mathcal{O}''_T$ operators are linear combinations of $\mathcal{O}_{S_L}$ and $\mathcal{O}_T$. Therefore, we get some non-zero value of this TPA for these two solutions. For the $\mathcal{O}'_{S_L}$ solution, $\langle A^{(2)}_T(q^2)\rangle$ reaches a maximum value of $\sim 0.7\%$ at $q^2 \simeq 6$ GeV$^2$ and decreases to zero at $q^2_{\rm max}$. For the $\mathcal{O}''_T$ solution, $\langle A^{(2)}_T(q^2)\rangle$ reaches a maximum value of $\sim 1.7\%$ at $q^2 \simeq 5.4$ GeV$^2$ and decreases to zero at $q^2_{\rm max}$.
\end{itemize} 
\begin{figure*}[ht] 
\centering
\begin{tabular}{cc}
\includegraphics[width=75mm]{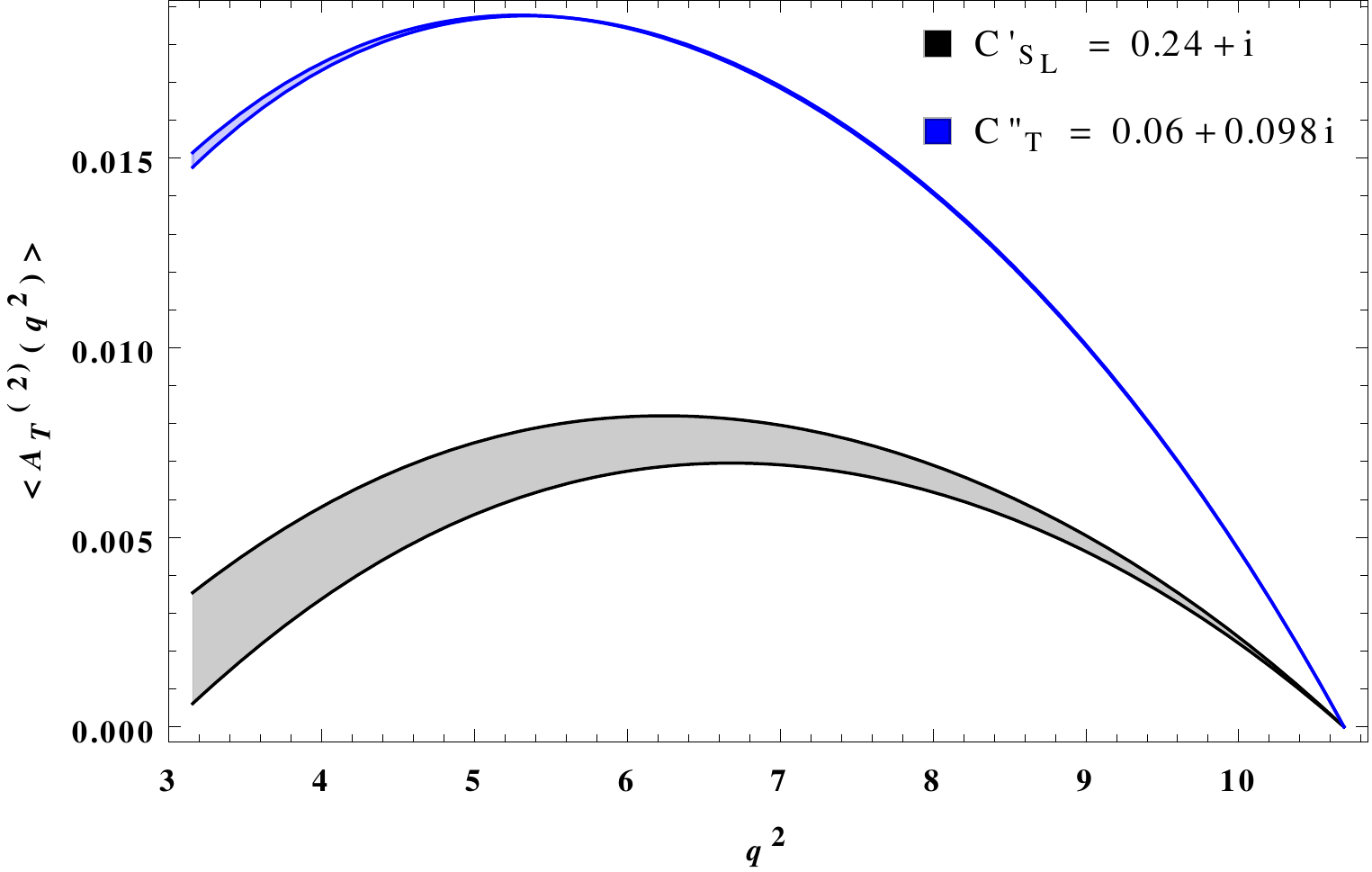}&
\includegraphics[width=75mm]{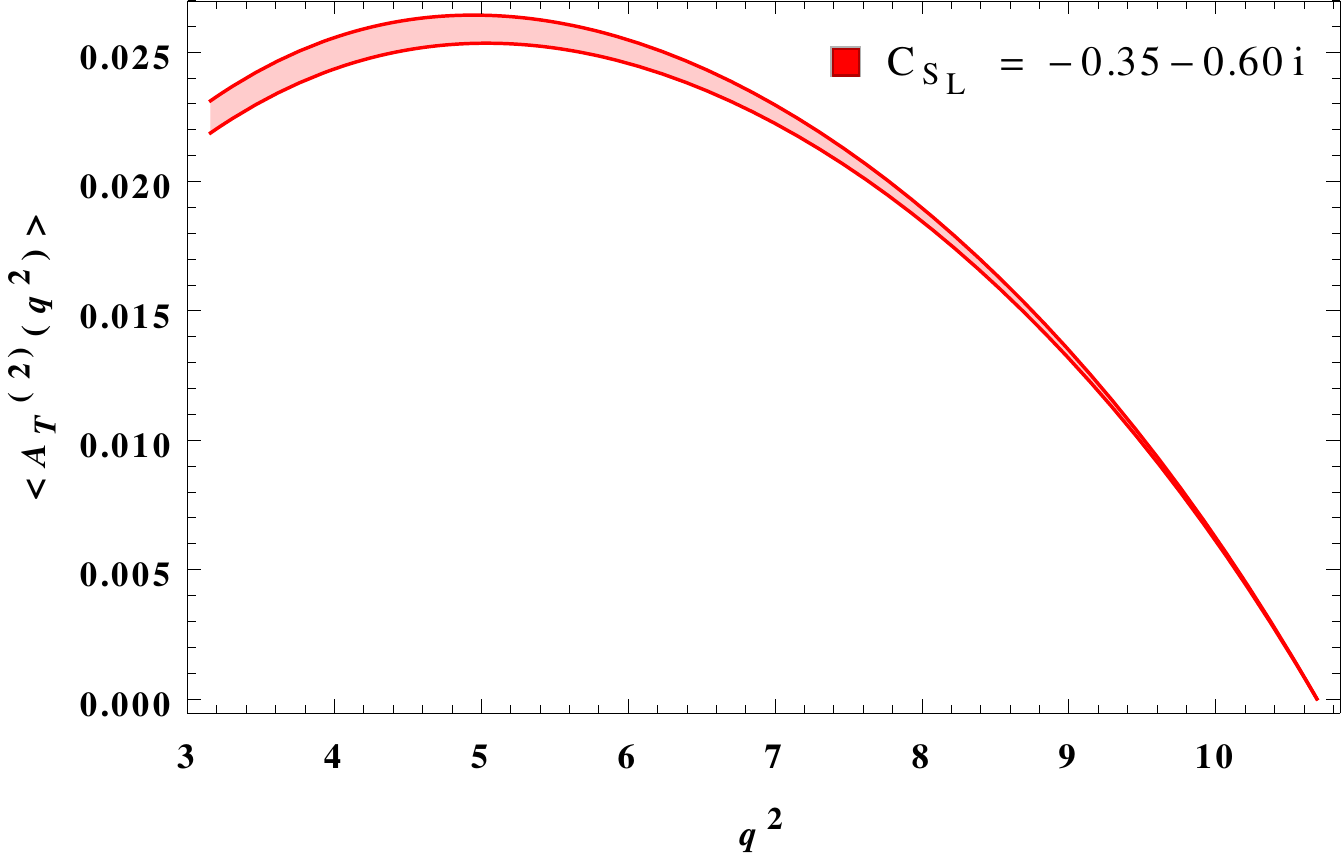}\\
\end{tabular}
\caption{The second TPA is plotted as a function of $q^2$ (GeV$^2$) for three benchmark NP WCs $C'_{S_L} = 0.24+i$ (blue curve), $C''_{T}= 0.06+ 0.098i$ (black curve) and $C_{S_L} = -0.35-0.60i$ (red curve).}
\label{fig5}
\end{figure*}
Our next aim is to compute the maximum CP violation allowed by the present $b\to c\tau\bar{\nu}$  data. To calculate this, we choose a benchmark point from the $1\sigma$ allowed parameter space of each NP solution. From Fig.~\ref{fig4}, we have learned that for any complex value of $C_{V_L}$ three TPAs lead to zero. Only the second TPA  $\langle A^{(2)}_T(q^2)\rangle$ is non-zero for the $\mathcal{O}'_{S_L}$ and $\mathcal{O}''_T$ solutions. Therefore, we pick a benchmark points from Fig~\ref{fig1} for each of these two solutions. These points are $C'_{S_L}= 0.24\pm i$ and $C''_T= 0.06+0.098i$ , which can lead to the maximum value of the TPA $\langle A^{(2)}_T(q^2)\rangle$ in $B\to D^*\tau\bar{\nu}$ decay. In the left panel of Fig.~\ref{fig5}, we plot the TPA $\langle A^{(2)}_T(q^2)\rangle$ as a function of $q^2$ for these two benchmark points of $\mathcal{O}'_{S_L}$ and $\mathcal{O}''_T$ solutions. From this plot, we observe that it has almost same features which are obtained from the plot of $\langle A^{(2)}_T(q^2)\rangle$ in Fig~\ref{fig4}. We have not got much larger value of TPA $\langle A^{(2)}_T(q^2)\rangle$ than what we got for the best fit NP solutions.

As per discussion in Sec II, the $\mathcal{O}_{S_L}$ solution listed in Table~\ref{tab1} is marginally disfavored because the best fit values of $C_{S_L}$ does not satisfy the constraint of $Br(B_c\to \tau\bar{\nu})<60\%$. However, a small fraction of the $1\sigma$ region of this solution falls on the region spanned by the constraint $30\%<Br(B_c\to \tau\bar{\nu})<60\%$.  For completeness, we calculate the predictions of TPAs for this solution. We can get a allowed value of $C_{S_L}$ which can give to maximum possible TPA for the $\langle A^{(2)}_T(q^2)\rangle$. We choose a benchmark point $C_{S_L} = -0.35-0.60i$ from the allowed region and calculate the second TPA. In right panel of Fig.~\ref{fig5}, we plot $\langle A^{(2)}_T(q^2)\rangle$ as a function of $q^2$ for the benchmark point of $C_{S_L}$. From this plot, we observe that the second TPA reaches a maximum value of $\sim 2.6\%$ at $q^2\simeq 5$ GeV$^2$ and decreases to zero at $q^2_{\rm max}$. In fact, this is the maximum value of $\langle A^{(2)}_T(q^2)\rangle$ predicted by the scalar operator solution among all the predictions made by allowed NP solutions.

\section{Conclusions}

In this work, we have done a global fit of $b\to c\tau\bar{\nu}$ data assuming NP WCs to complex. We find that the $\mathcal{O}_{V_L}$ solution is the only NP solution allowed by the constraint $Br(B_c\to \tau\bar{\nu})<10\%$. If we relax the constraint to $30\%$ or $60\%$, then we get one or two additional allowed NP solutions. We calculate the predictions of angular observables in $B\to (D,D^*)\tau\bar{\nu}$ decays. We find that the forward-backward asymmetries in these two decays are quite useful to distinguish the two solutions other than the $\mathcal{O}_{V_L}$ solution.

We then compute the maximum values of CP violating TPAs in  $B\to D^*\tau\bar{\nu}$ decay for the allowed NP solutions. These TPAs are zero in the SM. Hence any non-zero measurement of these quantities would give a smoking gun signal of physics beyond SM. Here we find that the predictions of first and third TPAs are zero for all NP solutions whereas the second TPA reaches a maximum value of $\sim 1.9\%$ for the $\mathcal{O}'_{S_L}$ solution and $\sim 0.9\%$ for the $\mathcal{O}''_T$ solution. The mildly favored NP solution $\mathcal{O}_{S_L}$ predicts a maximum value of $\sim 2.6\%$ for the second TPA which is the maximum predicted value among all the NP predictions.

To measure the angular observables and TPAs, the reconstruction of the $\tau$ lepton momentum is crucial. This is quite difficult because of the missing neutrinos. The LHCb collaboration has already made a fair attempt to reconstruct the $\tau$ lepton through $\tau^- \to \pi^-\pi^+\pi^-(n\pi^0)\nu_{\tau}$ decay channel~\cite{Aaij:2017uff}. However, in case of Belle II, it is very hard to reconstruct the $\tau$ momentum through leptonic decay $\tau^-\to \ell^- \nu_{\tau}\bar{\nu_{\ell}}$ because of multiple neutrinos in the final state. Thus, LHCb may be able to measure $\theta_{\tau}$ and $\phi$ with a better precision than Belle II and this could lead to a null test of the TPAs. We hope LHCb would be able to overcome this challenge in the near future~\cite{Cerri:2018ypt}. Recently in Ref.~\cite{Marangotto:2018pbs}, the author discussed an outline to measure the full angular distribution and the CP violating TPAs for $B\to D^*\ell\bar{\nu}$ decays at the collider experiments. 

\section*{Acknowledgements}
We would like to thank the organizers of WHEPP 2019 at IIT Guwahati, where this work had been initiated. We thank Amarjit Soni for useful suggestions at WHEPP. We also thank S. Uma Sankar for useful discussions and for  careful reading of the manuscript. 

\appendix
\section{Angular Coefficients}
\label{app}
The total longitudinal and transverse amplitudes are defined as~\cite{Duraisamy:2014sna}
\begin{equation}
A_L = \left(V_1^0 -\frac{1}{3} V_2^0\right), \quad A_T =2\left(V_1^T - \frac{1}{3}V_2^T \right).	
\end{equation}
The longitudinal coefficients $V^0_1$ and $V^0_2$ are written as
\begin{eqnarray}
V_1^0 &=& 2\left[\left(1+\frac{m^2_{\tau}}{q^2}\right)\left(|\mathcal{A}_0|^2 + 16|\mathcal{A}_{0T}|^2\right)+ \frac{2m^2_{\tau}}{q^2}|\mathcal{A}_{tP}|^2 -\frac{16m_{\tau}}{\sqrt{q^2}} {\rm Re}\left[\mathcal{A}_{0T}\mathcal{A}^*_0\right] \right], \nonumber \\
V_2^0 & =& 2\left(1-\frac{m^2_{\tau}}{q^2}\right)\left[-|\mathcal{A}_0|^2 + 16|\mathcal{A}_{0T}|^2\right],
\end{eqnarray}
and the transverse coefficients $V^T_1$, $V^T_2$ and $V^T_5$ are given by
\begin{eqnarray}
V_1^T &=& \frac{1}{2}\left(3+\frac{m^2_{\tau}}{q^2}\right)\left(|\mathcal{A}_{\parallel}|^2 + |\mathcal{A}_{\perp}|^2\right) + 8 \left(1+ \frac{3m^2_{\tau}}{q^2}\right)\left(|\mathcal{A}_{\parallel T}|^2 + |\mathcal{A}_{\perp T}|^2\right) - \frac{16m^2_{\tau}}{\sqrt{q^2}} {\rm Re} \left[\mathcal{A}_{\parallel T}\mathcal{A}^*_{\parallel}+ \mathcal{A}_{\perp T} \mathcal{A}^*_{\perp}\right], \nonumber \\
V_2^T & =& \left(1-\frac{m^2_{\tau}}{q^2}\right)\left[\frac{1}{2} \left(|\mathcal{A}_{\parallel}|^2 + |\mathcal{A}_{\perp}|^2\right) - 8 \left(|\mathcal{A}_{\parallel T}|^2 + |\mathcal{A}_{\perp T}|^2\right) \right], \nonumber \\
V_5^T &=& 2\left(1-\frac{m^2_{\tau}}{q^2}\right) {\rm Im} \left[\mathcal{A}_{\parallel} \mathcal{A}^*_{\perp} \right].
\end{eqnarray}
The expressions for mixed angular coefficients $V^{0T}_3$ and $V^{0T}_4$ are given by 
\begin{eqnarray}
V^{0T}_3 & =& 2\sqrt{2} \,{\rm Im} \left[-\mathcal{A}_{\parallel}\mathcal{A}^*_0 + \frac{m^2_{\tau}}{q^2} \mathcal{A}_{\perp} \mathcal{A}^*_{tP} +\frac{4m_{\tau}}{\sqrt{q^2}} \left(\mathcal{A}_{0T}\mathcal{A}^*_{\parallel} - \mathcal{A}_{\parallel T} \mathcal{A}^*_0 + \mathcal{A}_{\perp T} \mathcal{A}^*_{tP}\right) \right] , \nonumber \\
V^{0T}_4 & = & \sqrt{2} \left(1- \frac{m^2_{\tau}}{q^2} \, { \rm Im} \left[\mathcal{A}_{\perp} \mathcal{A}^*_0\right]\right).
\end{eqnarray}
The corresponding hadronics matrix elements are expressed as
\begin{eqnarray}
\mathcal{A}_0 & =& \frac{m_B+m_{D^*}}{2m_{D^*}\sqrt{q^2}}\left[\left(m^2_B - m^2_{D^*} -q^2\right)A_1(q^2) - \frac{\lambda_{D^*}}{(m_B+m_{D^*})^2}A_2(q^2)\right] (1+C_{V_L} -C_{V_R}), \nonumber \\
\mathcal{A}_{\pm} &=& (m_B+ m_{D^*})A_1(q^2)\left(1+C_{V_L}-C_{V_R}\right) \mp \frac{\sqrt{\lambda_{D^*}}}{(m_B+m_{D^*})} V(q^2)\left(1+ C_{V_L} + C_{V_R}\right), \nonumber \\
\mathcal{A}_t & =& \frac{\sqrt{\lambda_{D^*}}}{\sqrt{q^2}} A_0(q^2) (1+C_{V_L} -C_{V_R}), \nonumber \\
\mathcal{A}_P &=& \frac{\sqrt{\lambda_{D^*}}}{m_b +m_c} A_0(q^2) (C_{S_R} -C_{S_L}), \nonumber \\
\mathcal{A}_{0T} & =&  \frac{C_T}{2m_{D^*}}\left[\left(m^2_B + 3 m^2_{D^*} -q^2\right) T_2(q^2) - \frac{\lambda_{D^*}}{m^2_B-m^2_{D^*}}T_3(q^2)\right], \nonumber \\
\mathcal{A}_{\pm T} &= & C_T\left[\frac{m^2_B -m^2_{D^*}}{\sqrt{q^2}}T_2(q^2) \pm \sqrt{\frac{\lambda_{D^*}}{q^2}}T_1(q^2)\right].
\end{eqnarray} 
Further the transversity amplitudes can be defined as
\begin{equation}
\mathcal{A}_{\parallel(T)} = \frac{1}{\sqrt{2}}\left(\mathcal{A}_{+(+T)} + \mathcal{A}_{-(-T)}\right), \quad \mathcal{A}_{\perp (T)}  = \frac{1}{\sqrt{2}}\left(\mathcal{A}_{+(+T)} - \mathcal{A}_{-(-T)}\right).	 
\end{equation}
The amplitude $\mathcal{A}_{tP}$ is a combination of $t$ and $P$ amplitudes which is given by 
\begin{equation}
\mathcal{A}_{tP} =  \left(A_t + \frac{\sqrt{q^2}}{m_{\tau}}\mathcal{A}_P\right)
\end{equation} 
All the above expressions for the angular coefficients and hadronic amplitudes are taken from the Ref.~\cite{Duraisamy:2014sna}. The form factors appeared in the hadronic amplitudes $V(q^2)$, $A_{0,1,2}(q^2)$ and $T_{1,2,3}(q^2)$ are calculated in HQET parametrization~\cite{Caprini:1997mu} and their expressions can be also found in Ref.~\cite{Sakaki:2013bfa}.


\end{document}